\documentclass[floafix,prd,showpacs,showkeys,preprintnumbers,nofootinbib,superscriptaddress,11pt]{revtex4-1}
\usepackage[utf8]{inputenc}
\usepackage[sort&compress]{natbib}
\usepackage{ulem}
\usepackage{bm}
\usepackage{times}
\usepackage{amssymb,amsbsy,amsmath,amsfonts}
\usepackage{graphicx}
\usepackage{epstopdf}
\usepackage{float}
\usepackage{color}
\usepackage{morefloats}
\usepackage{rotating}
\usepackage{srcltx}
\usepackage{slashed}
\usepackage{subfigure}
\usepackage{multirow}
\usepackage{verbatim}
\usepackage{hyperref}
\usepackage{lineno}
\usepackage{overpic}
\usepackage{booktabs}
\usepackage{tabularx}

\begin{document}
%\linenumbers

\title{ Exploring  the nature of $Y(4230)$ and $Y(4360)$  in B decays }

\author{Ming-Zhu Liu}
\affiliation{
Frontiers Science Center for Rare isotopes, Lanzhou University,
Lanzhou 730000, China}
\affiliation{ School of Nuclear Science and Technology, Lanzhou University, Lanzhou 730000, China}

\author{Qi Wu}
\email[Corresponding author:]{wuqi@htu.edu.cn}
\affiliation{Institute of Particle and Nuclear Physics, Henan Normal University, Xinxiang 453007, China}

\begin{abstract}

Vector charmonium states can be directly produced  in the $e^+e^{-}$ annihilation process.  Among them, $Y(4230)$ and $Y(4360)$ splitting from the previously   discovered   $Y(4260)$ are not easily arranged into the conventional charmonium spectrum, while recent studies have indicated that they  have strong couplings to  $D\bar{D}_1$ and $D^*\bar{D}_1$. In this work, we investigate  the production of  $Y(4230)$ and $Y(4360)$  as the heavy-quark spin symmetry doublet hadronic molecules of  $D\bar{D}_1$ and $D^*\bar{D}_1$  in  $B$ decays via  the triangle diagram mechanism. In particular, we propose that   the  decay  constants of  $Y(4230)$ and $Y(4360)$ extracted  in  $B$ decays are useful for clarifying  their nature.

\end{abstract}

%\pacs{13.60.Le, 12.39.Mk,13.25.Jx}

\maketitle

\section{Introduction}

The charmonium spectrum plays an important role in  understanding strong interactions as well as in developing effective field theories(EFTs) and quantum chromodynamics (QCD)-inspired  potential models, where the former consists of non-relativistic QCD (NRQCD)~\cite{Caswell:1985ui} and potential NRQCD~\cite{Pineda:1997bj}, and the latter includes the Cornell potential model~\cite{Eichten:1974af}, the Godfrey-Isgur(GI) model~\cite{Godfrey:1985xj} and other improved potential models~\cite{Li:2009zu,Ortega:2009hj,Tan:2019qwe,Duan:2020tsx,Deng:2023mza}.  The potential between quark and anti-quark can be obtained from lattice QCD~\cite{Bali:2005fu} and the QCD-inspired  potential model. Most potential  models can effectively describe the mass spectrum of charmonium states below the $D\bar{D}$ mass threshold, but face challenges when dealing with charmonium states in the vicinity  of the mass thresholds of  a pair of charmed mesons~\cite{Barnes:2005pb}. The states beyond the conventional quark model are referred to as $XYZ$ states, motivating intensive discussions on their nature (see some recent reviews~\cite{Chen:2016qju,Lebed:2016hpi,Oset:2016lyh,Esposito:2016noz,Dong:2017gaw,Guo:2017jvc,Olsen:2017bmm,Karliner:2017qhf,Brambilla:2019esw,Guo:2019twa,Yang:2020atz,Meng:2022ozq,Liu:2024uxn}).

The X(3872) discovered by the Belle Collaboration opens a new era of charmonium  research~\cite{Belle:2003nnu}, since its mass is lower than the mass of the conventional  $\chi_{c1}(2P)$ state, predicted by the  legendary GI model,  by $90$~MeV~\cite{Godfrey:1985xj}. Meanwhile, the $X(3915)$ and $X_2(3930)$ reported  in the Review of Particle Physics(RPP)~\cite{ParticleDataGroup:2022pth} are not easily assigned as conventional  $\chi_{c0}(2P)$ and $\chi_{c2}(2P)$ states~\cite{Guo:2012tv,Olsen:2014maa}. Taking into account the effect of a pair of charmed mesons greatly improves our ability to address their mass puzzles~\cite{Li:2009zu,Ortega:2009hj,Tan:2019qwe,Duan:2020tsx,Deng:2023mza}, indicating that these states have strong couplings to a pair of charmed mesons to the extent that they are regarded as hadronic molecules. From analysis of the invariant mass distribution of $X(3872)$, the BESIII and LHCb collaborations found that the molecular component $D\bar{D}^*$ accounts  for more than $80\%$ of its total wave function~\cite{BESIII:2023hml,LHCb:2020xds}.

For vector charmonium states, the $Y(4260)$ was firstly  observed by the BaBar Collaboration in  the  process of  $e^+e^-\to\pi^+\pi^-J/\psi$~\cite{BaBar:2005hhc},  and later  confirmed by the CLEO Collaboration~\cite{CLEO:2006ike} and the Belle Collaboration~\cite{Belle:2007dxy}. The peak energy region around  4260 MeV was  not pronounced from  the measured  $R$ value of the  BESIII Collaboration~\cite{BES:2007zwq}, which implies that there likely  exists a state beyond the conventional vector charmonia~\cite{Chen:2016qju}. With more precise data samples in the $e^+e^-\to\pi^+\pi^-J/\psi$ process, the BESIII Collaboration found that the original $Y(4260)$ state splits into two states, $Y(4220)$ and $Y(4320)$~\cite{BESIII:2016bnd}. Two fine structures were later confirmed  in the processes of   $e^+e^-\to\pi^+\pi^-h_c$~\cite{BESIII:2016adj},   $e^+e^-\to\pi^+\pi^-\psi(2S)$~\cite{BESIII:2017tqk,BESIII:2021njb}, and $e^+e^-\to J/\psi \eta$~\cite{BESIII:2020bgb}, with average  masses and widths of $(4222.5\pm 2.4, 48 \pm 8)$ MeV and  $(4374 \pm 7, 118\pm 12)$ MeV, denoted by $Y(4230)$ and $Y(4360)$.   However,  they were not  arranged into the conventional  charmonium spectrum~\cite{Deng:2023mza},  which has motivated much discussion on their properties and internal structures.

In Ref.~\cite{Chen:2015bft}, Chen et al. assigned $Y(4260)$ and $Y(4360)$ as fake states via Fano-like interference phenomena. Taking into account the $S-D$ mixing effect, the authors argued that $Y(4230)$ and $Y(4360)$ are the  mixing states of   $\psi(4S)$ and $\psi(3D)$~\cite{Fu:2018yxq,Wang:2019mhs}. In Ref.~\cite{Wang:2021qus}, Wang assigned  the  $Y(4230)$ and $Y(4360)$  as compact tetraquark states. In Ref.~\cite{Zhou:2023yjv}, Zhou et al. simulated the cross sections of $e^+e^-$ to several open charm channels, indicating that $Y(4230)$ and $\psi(4160)$ are the same  $\psi (2 ^3D_1)$ state with the pole position (4222, 32) MeV, and that the enhancement at energy of $4160$ MeV may be attributed to the interference effect of   the  bare $c\bar{c}$ state and a pair of charmed mesons. After simulating additional data samples, Nakamura et al. argued that the $\psi(4160)$ does not exist, but there exist two poles at energy $4230$~MeV~\cite{Nakamura:2023obk}.  Detten et al. simulated the data samples in  the energy region from $4.2$ to $4.35$ GeV, assigning the $Y(4230)$ as the bound state of $\bar{D}D_1$~\cite{vonDetten:2024eie}.  Due to the $Y(4230)$ and $Y(4360)$ in the vicinity of $\bar{D}D_1$ and $\bar{D}^*D_1$ mass thresholds, as well as their mass splitting equivalent to the $\pi$ mass, $Y(4230)$ and $Y(4360)$ are expected as the bound states of $\bar{D}D_1$ and $\bar{D}^* D_1$, respectively. In Refs.~\cite{Wang:2016wwe,Anwar:2021dmg,Peng:2022nrj,Ji:2022blw,Wang:2023ivd,Liu:2024ziu}, the authors assigned $Y(4230)$ and $Y(4360)$  as the heavy quark spin symmetry (HQSS) doublet hadronic molecules of $\bar{D}^{(*)}D_1$. At this time, discussions on the nature of $Y(4230)$ and $Y(4360)$ are still  ongoing.

In addition to the mass spectra of $Y(4230)$ and $Y(4360)$, their decay behaviors and  production mechanisms offer important ways to probe their properties. Identifying $Y(4230)$  and $Y(4360)$ as the bound states of $\bar{D}D_1$ and  $\bar{D}^*D_1$, respectively, we predicted  their pionic decay and radiative decay widths~\cite{Liu:2024ziu}. In Refs.~\cite{Wang:2013cya,Dong:2013kta,Liu:2013vfa,Guo:2013zbw,Dong:2014zka,Qin:2016spb,Chen:2017abq}, assuming that $Y(4260)$ and $Y(4390)$ are bound states of $\bar{D}D_1$ and  $\bar{D}^*D_1$, their different decay modes were investigated. In Ref.~\cite{Wang:2023dsm}, assuming the $Y(4220)$ as a compact tetraquark state, its  three-body decay widths were predicted. According to RPP, $J/\psi$, $\psi(3686)$, $\psi(4040)$, and $\psi(4415)$ as the conventional    $\psi(1S)$,  $\psi(2S)$, $\psi(3S)$, and $\psi(4S)$ states and $\psi(3770)$ and $\psi(4160)$ as the conventional  $\psi(1D)$ and   $\psi(2D)$ states have  been  fully  observed in $B$ decays. However, the Belle Collaboration and LHCb Collaboration did not observe a significant signal of  $Y(4260)$ in the $B$ decay~\cite{BaBar:2005xmz,Belle:2019pfg,LHCb:2022oqs}. The observation of $Y(4230)$ and $Y(4360)$ in $B$ decays and the understanding of the related production mechanisms have important implications for probing their inner structure. As a result, it is necessary to explore the production of $Y(4230)$ and $Y(4360)$ in $B$ decays.

The production of $Y(4260)$ as a mixture of the charmonium state and the compact tetraquark state  in $B$ decays was explored using the QCD sum rule~\cite{Albuquerque:2015nwa}. In the present work, identifying $Y(4230)$ and $Y(4360)$ as the HQSS  doublet hadronic molecules of $\bar{D}^{(*)}D_1$, we adopt the triangle diagram to investigate the production of $Y(4230)$ and $Y(4360)$ in $B$ decays.   Since the topological diagrams illustrating  the  production of  hadronic molecules in heavy hadron decays cannot be factorized~\cite{Chen:2020eyu}, final-state interaction is one effective approach for dealing  with the non-factorized topological diagrams; this method has been widely applied in heavy flavor physics~\cite{Cheng:2004ru,Yu:2017zst} and the production of exotic states in heavy hadron decays~\cite{Liu:2006df,Wu:2019rog,Chen:2020eyu,Liu:2020orv,Liu:2022dmm,Liu:2024uxn}.

This paper is organized as follows. After the Introduction, we introduce the production of $Y(4230)$ and $Y(4360)$ in B decays via the triangle mechanism and present the relevant Lagrangians in Sect.~\ref{sec:Sec2}. The numerical results and discussion for the branching fractions of $\mathcal{B}[B \rightarrow K Y(4230)]$ and  $\mathcal{B}[B \rightarrow K Y(4360)]$   and the decay constants of $Y(4230)$ and $Y(4360)$  are shown  in Sect.~\ref{Sec:Num}, and a brief summary is given in Sect.~\ref{sec:summary}.

\section{Theoretical formalism}
\label{sec:Sec2}

\begin{figure}[htb!]
\begin{tabular}{ccc}
  \centering
\includegraphics[width=0.9\textwidth, trim=0.2cm 10cm 0.2cm 10cm, clip]{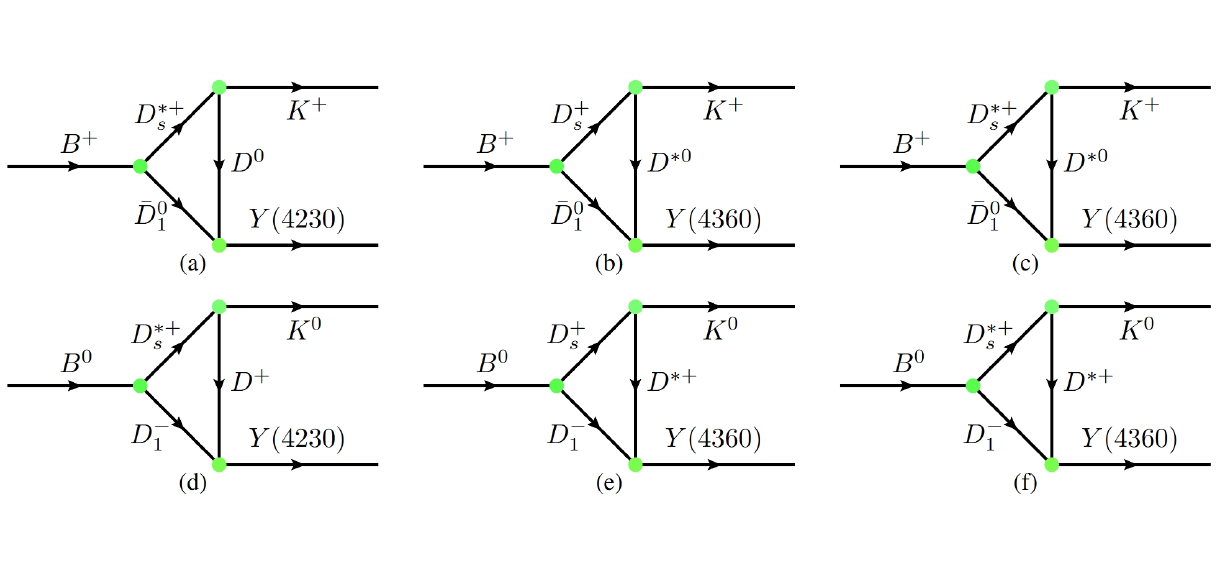}
 \end{tabular}
  \caption{Triangle diagrams accounting for the weak decays of $B^+ \rightarrow K^+ Y(4230)$~(\textbf{a}), $B^+ \rightarrow K^+ Y(4360)$ (\textbf{b, c}), $B^0 \rightarrow K^0 Y(4230)$~(\textbf{d}),  and $B^0 \rightarrow K^0 Y(4360)$ (\textbf{e, f}).}\label{Fig:2}
\end{figure}

In this work, the production of $Y(4230)$ and $Y(4360)$ as the bound states of $\bar{D}D_1$ and $\bar{D}^*D_1$ in $B$ decays is illustrated  by the triangle diagrams as shown in Fig.~\ref{Fig:2}, where the $B$ meson weakly decays into $D_{s}^{(*)}\bar{D}_1$ then the $D_{s}^{(*)}$ mesons scatter into $D^{(*)}$ and $K$ mesons, and finally $Y(4230)$ and $Y(4360)$ are dynamically generated via the $\bar{D}_1 D$ and $\bar{D}_1 D^*$ interactions.  Here, we employ  the effective Lagrangian approach to calculate the Feynman diagrams of Fig.~\ref{Fig:2}. The relevant Lagrangians describing the interactions of each vertex in the triangle diagram are presented in the following.

The weak decays $B^+ \to   \bar{D}_1^0 D_s^{(*)+}$ mainly  proceed via the external $W$-emission mechanism at the quark level, which always play a dominant  role in terms of the topological  classification of weak decays~\cite{Chau:1987tk,Ali:1998eb,Ali:2007ff,Li:2012cfa}.
In the naive factorization ansatz~\cite{Bauer:1986bm}, the amplitudes of the weak decays $B^+ \to   \bar{D}_1^0 D_s^{(*)+}  $ can be expressed as the products of two current matrix elements
\begin{eqnarray}\label{Ds-KK}
\mathcal{A}\left(B \to D_s \bar{D}_1 \right)&=&\frac{G_{F}}{\sqrt{2}} V_{cb}V_{cs} a_{1}\left\langle D_s|J_{\mu}| 0\right\rangle\left\langle \bar{D}_1|J^{\mu}| B\right\rangle, \\
\mathcal{A}\left(B \to D^{*}_s \bar{D}_1 \right)&=&\frac{G_{F}}{\sqrt{2}} V_{cb}V_{cs} a_{1}\left\langle D_s^{*}|J_{\mu}| 0\right\rangle\left\langle \bar{D}_1|J^{\mu}| B\right\rangle,
\end{eqnarray}
where $J_\mu=\bar{q}_1 \gamma_\mu(1-\gamma_5)q_2$,  $G_F$ is the Fermi constant, $V_{cb}$ and $V_{cs}$ are  the Cabibbo-Kobayashi-Maskawa (CKM) matrix elements, and $a_{1}$ is the  Wilson coefficient dependent on the renormalization scale~\cite{Chau:1987tk,Cheng:1993gf,Cheng:2010ry}. In this work, we deal with $B$ decays, and therefore extract the value of the effective Wilson coefficient at the energy scale $\mu=m_b$, e.g., $a_1=1.07$~\cite{Li:2012cfa}, consistent with Ref.~\cite{Lu:2009cm}. The former current matrix elements  are written as
\begin{eqnarray}
\langle0|J_\mu|D_s(p_1)\rangle&=&-i f_{D_s} p_{1\mu},  \nonumber \\
\langle0|J_\mu|D_{s}^*(p_1,\epsilon)\rangle&=&f_{D_{s}^*} \epsilon_\mu m_{D_{s}^*},
\label{Eq:ME1}
\end{eqnarray}
where  $f_{D_s}$ and  $f_{D_s^*}$ are   the  decay constants of $D_{s}$ and $D_{s}^*$. In this work, we take  $G_F = 1.166 \times 10^{-5}~{\rm GeV}^{-2}$, $V_{cb}=0.041$,   $V_{cs}=0.987$, $a_1=1.07$, $f_{D_{s}} = 250$ MeV, and $f_{D_{s}^{\ast}}=272$~MeV~\cite{ParticleDataGroup:2022pth,Verma:2011yw,FlavourLatticeAveragingGroup:2019iem,Li:2017mlw,Li:2012cfa}.   The  latter current matrix element describing the transition of $B(k_0)\to D_1(q_2)$ is characterized by  a series of  form factors
\begin{eqnarray}
&&\langle D_1| J^{\mu}|B\rangle =i\varepsilon^{\mu\nu\alpha\beta}\epsilon_\nu P_\alpha q_\beta \frac{A(q^2)}{m_B-m_{D_1}} +\varepsilon^{\mu} (m_B-m_{D_1})V_1(q^2) \nonumber \\
&&-\varepsilon\cdot P\left\{ P^{\mu} \frac{V_2(q^2)}{m_B-m_{D_1}}+ \frac{q^{\mu}}{q^2}\left[(m_B-m_{D_1})V_1(q^2)-(m_B+m_{D_1})V_2(q^2)-2m_{D_1}V_0(q^2)\right]\right\}\label{Eq:ME2}
\end{eqnarray}
with $A(q^2)$, $V_0(q^2)$, $V_1(q^2)$ and $V_2(q^2)$ being  the form factors, where $P=(k_0+q_2)$ and $q=(k_0-q_2)$,  and $\epsilon_\nu$ is the polarization vector of $D_1$. In general,  the form factors  are parameterized in the following form
\begin{eqnarray}
F(q^2)=\frac{F(0)}{1-a\zeta+b\zeta^2},\label{Eq:A1}
\end{eqnarray}
with  $\zeta=q^2/m^2_{B}$, where $F(0)$, $a$, and $b$ are parameters determined in phenomenological models. Herer, we take these parameters determined in the covariant light-front quark model~\cite{Verma:2011yw}, where the uncertainties of their values are given as around $1\%$.  In Table~\ref{BtoDformfactor}, we collect the values of  $F(0)$, $a$, and $b$ in the form factors to be used in this work.

 \begin{table}[ttt]
 \centering
 \caption{Values of  $F(0)$, $a$, and  $b$ in the $B \rightarrow D_1$ and $B \rightarrow K$ transition form factors~\cite{Verma:2011yw}\label{BtoDformfactor} }
 \begin{tabular}{ccccc|ccccccccc}
 \hline\hline
   &   $A$~~~ & $V_0$~~~ &  $V_1$~~~ & $V_2$~~~   &~~~~~~  & $F_0$~~~   & $F_1$~~~  \\
 \hline
 $F(0)^{B\to D_1}$&  0.25~~~  & 0.52~~~ & 0.58~~~ & -0.10~~~ & $F(0)^{B\to K}$   & 0.34~~~ & 0.34~~~ \\
 $a^{B\to D_1}$  & 1.17~~~ &  1.14~~~  & -0.25~~~ & -5.95~~   & $a^{B\to K}$ & 0.78~~~ & 1.60~~~\\
 $b^{B\to D_1}$ & 0.33~~~  & 0.34~~~  & 0.29~~~ &26.2~~~ & $b^{B\to K}$ & 0.05~~~ & 0.73~~~ \\
\hline\hline
 \end{tabular}
 \end{table}

With Eqs.(\ref{Eq:ME1})-(\ref{Eq:ME2}), we obtain amplitudes for the decays $B(k_0) \to D^{(*)}_s(q_1) \bar{D}_1(q_2)$,
\begin{linenomath*}
\begin{align}
%\mathcal{A}(B\to D_{s}\bar{D}_1)&= \frac{G_{F}}{\sqrt{2}}V_{cb}V_{cs} a_{1} f_{D_{s}}\epsilon^\mu(q_2)
%\Big(q_{1\mu} (m_B-m_{D_1})V_1(q^2)-(k_0+q_2)_\mu \\ \nonumber
%&\times\left[\frac{(k_0+q_2)\cdot q_1}{m_B-m_{D_1}} V_2(q^2)+ (m_B-m_{D_1})V_1(q^2)-(m_B+m_{D_1})V_2(q^2)-2m_{D_1}V_0(q^2)\right] \Big) \\ \nonumber
\mathcal{A}(B\to D_{s}\bar{D}_1)&= -i\frac{G_{F}}{\sqrt{2}}V_{cb}V_{cs} a_{1} f_{D_{s}}\epsilon^\mu(q_2)
\Big[-2q_{2\mu} (m_B-m_{D_1})V_1(q^2_1)+2m_{D_1}(k_0+q_2)_\mu V_0(q^2_1) \nonumber\\
& -\Big(\frac{(k_0+q_2)_\mu (k_0+q_2)\cdot q_1}{m_B-m_{D_1}}-(m_B+m_{D_1})(k_0+q_2)_\mu \Big)V_2(q^2_1)\Big], \\ \nonumber
\mathcal{A}(B\to D_{s}^{\ast}\bar{D}_1)&= \frac{G_{F}}{\sqrt{2}}V_{cb}V_{cs} a_{1} m_{D_{s}^{\ast}}f_{D_{s}^{\ast}}\epsilon^\rho(q_1)\epsilon^\sigma(q_2)
\Big[(g_{\rho\sigma}-\frac{(q_1)_\rho (k_0+q_2)_\sigma}{q^2_1})(m_B-m_{D_1})V_1(q^2_1)\\ \nonumber
&+\Big((m_B+m_{D_1})\frac{(q_1)_\rho (k_0+q_2)_\sigma}{q^2_1}-\frac{(k_0+q_2)_\rho (k_0+q_2)_\sigma}{m_B-m_{D_1}}\Big)V_2(q^2_1) \\
&+2m_{D_1}\frac{(q_1)_\rho (k_0+q_2)_\sigma}{q^2_1} V_0(q^2_1)+\frac{i}{m_B-m_{D_1}}\varepsilon_{\mu\nu\alpha\beta}g^\mu_\rho g^\nu_\sigma (k_0+q_2)^\alpha (q_1)^\beta A(q^2_1)\Big].
\end{align}
\end{linenomath*}
The form factors characterizing the $B \to \bar{D}_1$ transition are uncertain due to the $P$-wave charmed meson $\bar{D}_1$. In this work, we use the branching fraction of the decay $B\to \bar{D}_1 \pi$ to reduce the uncertainty. Assuming the decay  $B\to \bar{D}_1 \pi$ can be factorized, we obtain the branching fraction  $\mathcal{B} (B\to \bar{D}_1 \pi)=3.68\times 10^{-3}$ via the naive factorization approach, which is larger than the experimental data i.e., $\mathcal{B} (B\to \bar{D}_1 \pi)=(1.5\pm0.6)\times10^{-3}$. Since the naive factorization approach can deal well with Cabibbo-favored decays, we supplement a unified parameter $0.63$ to the above form factors to reduce uncertainty.

The Lagrangians  describing the interactions between charmed mesons and a kaon meson are~\cite{Azevedo:2003qh}
\begin{linenomath*}
\begin{eqnarray}
\mathcal{L}_{D_{s}^{\ast} D K}&=& -i g_{D_{s}^{\ast} D K} ( D \partial^{\mu} K D_{s\mu}^{\ast\dag}-  D_{s\mu}^{\ast}\partial^{\mu} K D^{\dag}),\nonumber \\
\mathcal{L}_{D_{s} D^{\ast} K}&=& -i g_{D_{s} D^{\ast} K} (D_{s} \partial^{\mu} K D^{\ast\dag}_{\mu}-D_{\mu}^* \partial^{\mu} K  D_{s}^{\dag}), \nonumber \\
\mathcal{L}_{D_{s}^{\ast} D^{\ast} K}&=& - g_{D_{s}^{\ast} D^{\ast} K} \varepsilon_{\mu\nu\alpha\beta} \partial^{\mu}D_{s}^{\ast\nu} {\partial}^{\alpha} {D}^{\ast\beta\dag} K ,
\end{eqnarray}
\end{linenomath*}
where $g_{D_{s} D^{\ast} K}$, $g_{D_{s}^{\ast} D K}$, and $ g_{D_{s}^{\ast} D^{\ast} K}$ are charmed mesons coupling to kaon meson. In this work, we take the values of these couplings
$g_{D_{s} D^{\ast} K}=g_{D_{s}^{\ast} D K}=10$ and  $g_{D_{s}^{\ast} D^{\ast} K}=7.0$~GeV$^{-1}$ from Ref.~\cite{Wu:2023rrp}.

The  Lagrangians describing  $Y$ states coupling  to  the corresponding constituents are written as
\begin{eqnarray}
{\cal L}_{Y(4230)\bar{D}_1 D} &=& g_{Y(4230)\bar{D}_1 D}Y^\mu(4230)\bar{D}_{1\mu}{D},\nonumber \\
{\cal L}_{Y(4360)\bar{D}_1 D^*} &=& g_{Y(4360)\bar{D}_1 D^*}\varepsilon^{\mu\nu\alpha\beta}\partial_\mu Y_\nu(4360)\bar{D}_{1\alpha}{D}^*_\beta,
\end{eqnarray}
where $g_{Y(4230)\bar{D}_1 D}$ and $g_{Y(4360)\bar{D}_1 D^*}$ are the couplings between the molecules and their constituents. Following Ref.~\cite{Liu:2024ziu},  we employ contact-range EFTs to construct the $\bar{D}D_1$ and $\bar{D}^*D_1$ potentials, and then the  couplings are estimated from the residues of the poles by solving the Lippmann-Schwinger equation. In the isospin limit,  we take the values of  $g_{Y(4230)\bar{D}_1^0 D^0}=g_{Y(4230){D}_1^- D^+}=\frac{31.32}{2}$ GeV and $g_{Y(4360)\bar{D}_1^0 D^{*0}}=g_{Y(4360){D}_1^- D^{*+}}=\frac{6.97}{2}$~\cite{Liu:2024ziu}.

With the above effective Lagrangians, we can obtain the amplitudes for   $B(p)\rightarrow D^{*}_s(q_1) \bar{D}_1(q_2) [D^{(*)}(q_3)]\rightarrow K(p_1) +Y(p_2)$ corresponding to diagrams Fig.~\ref{Fig:2}a-f
\begin{eqnarray}
\mathcal{M}_{a/d}&=&i^3 \int\frac{d^4 q_3}{(2\pi)^4}\Big[\mathcal{A}_{\rho\sigma}(B\to D_{s}^{\ast}\bar{D}_1)\Big]\Big[g_{D^\ast_s DK}p_{1\tau}\Big]\Big[g_{Y(4230)D_1 D}\epsilon^Y_\xi\Big]\nonumber\\
&&\frac{-g^{\rho\tau}+q^\rho_1 q^\tau_1 /q^2_1}{q^2_1-m^2_1}\frac{-g^{\sigma\xi}+q^\sigma_2 q^\xi_2 /q^2_2}{q^2_2-m^2_2}\frac{1}{q^2_3-m^2_3}\mathcal{F}(q^2_3,m_3^2),\\
\mathcal{M}_{b/e}&=&i^3 \int\frac{d^4 q_3}{(2\pi)^4}\Big[\mathcal{A}_{\sigma}(B\to D_{s}\bar{D}_1)\Big]\Big[[-g_{D_s D^*K}p_{1\delta}\Big]]\Big[ig_{Y(4360)D_1 D^*}\varepsilon_{\mu\nu\alpha\beta}p^\mu_2 \epsilon^\nu_Y\Big]\nonumber\\
&&\frac{1}{q^2_1-m^2_1}\frac{-g^{\sigma\alpha}+q^\sigma_2 q^\alpha_2 /q^2_2}{q^2_2-m^2_2}\frac{-g^{\delta\beta}+q^\delta_3 q^\beta_3 /q^2_3}{q^2_3-m^2_3}\mathcal{F}(q^2_3,m_3^2),\\
\mathcal{M}_{c/f}&=&i^3 \int\frac{d^4 q_3}{(2\pi)^4}\Big[\mathcal{A}_{\rho\sigma}(B\to D_{s}^{\ast}\bar{D}_1)\Big]\Big[g_{D^\ast_s D^*K}\varepsilon_{\delta\kappa\zeta\theta}q^\delta_1 q_3^\zeta\Big]\Big[ig_{Y(4360)D_1D^*}\varepsilon_{\mu\nu\alpha\beta}p^\mu_2 \epsilon^\nu_Y\Big]\nonumber\\
&&\frac{-g^{\rho\kappa}+q^\rho_1 q^\kappa_1 /q^2_1}{q^2_1-m^2_1}\frac{-g^{\sigma\alpha}+q^\sigma_2 q^\alpha_2 /q^2_2}{q^2_2-m^2_2}\frac{-g^{\theta\beta}+q^\theta_3 q^\beta_3 /q^2_3}{q^2_3-m^2_3}\mathcal{F}(q^2_3,m_3^2),
\end{eqnarray}
%where $\mathcal{P}^1_{\mu\nu}=-\tilde{g}_{\mu\nu}$ and $\mathcal{P}^2_{\mu\nu\mu^\prime\nu^\prime}=\frac{1}{2}(\tilde{g}_{\mu\mu^\prime}\tilde{g}_{\nu\nu^\prime}+\tilde{g}_{\mu\nu^\prime}\tilde{g}_{\nu\mu^\prime})-\frac{1}{3}\tilde{g}_{\mu\nu}\tilde{g}_{\mu^\prime\nu^\prime}$ are the
%projection operator of the particle with spin 1 and 2, respectively. And $-\tilde{g}_{\mu\nu}(p)=-g_{\mu\nu}+\frac{p_\mu p_\nu}{p^2}$.
where $\epsilon_Y$  denotes the polarization vector of the $Y$  state.  In addition, to   eliminate the ultraviolet divergence of the loop functions and take into account the off-shell effect of exchanged mesons, we supplement  a dipole form factor
\begin{eqnarray}
\mathcal{F}\left(q^{2}, m^2\right)=\Big(\frac{m^2-\Lambda^{2}}{q^{2}-\Lambda^{2}}\Big)^2,\label{Eq:FFs1}
\end{eqnarray}
where the parameter $\Lambda$ can be further parameterized as $\Lambda=m+\alpha\Lambda_{\rm QCD} $ with $\Lambda_{\rm QCD}=0.22 \ {\rm GeV}$~\cite{Cheng:2004ru},  and $m$ is the mass of the exchanged meson.

With  the amplitudes  for the weak  decays  $B \to Y(4230) K$ and $B \to Y(4360) K$ given above, one can compute their partial decay widths
 %\begin{linenomath*}
 \begin{eqnarray}
\Gamma=\frac{1}{2J+1}\frac{1}{8\pi}\frac{|\vec{p}|}{m_{B}^2}{|\overline{M}|}^{2},
\end{eqnarray}
%\end{linenomath*}
where $J=0$ is the total angular momentum of the initial $B$ meson, the overline indicates the sum over the polarization vectors of the final states, and $|\vec{p}|$ is the momentum of either final state in the rest frame of the $B$ meson.

\section{Numerical results and discussion}
\label{Sec:Num}

In Table~\ref{mass},  we tabulate the masses and quantum numbers of relevant particles. One can see that there exists an unknown parameter $\alpha$ in our model, which cannot be determined from first principles.  Referring to previous studies of hadron-hadron interactions~\cite{He:2019rva,Wu:2021udi,Yu:2017zst,Wu:2019rog,Chen:2020eyu}, the parameter $\alpha$ is assigned values from $1$ to $5$ to show the uncertainties of our results.

\begin{table}[ttt]
\caption{Masses and quantum numbers of relevant hadrons needed in this work~\cite{ParticleDataGroup:2022pth} \label{mass}}
\begin{tabular}{ccc|cccccc}
  \hline\hline
   Hadron & $I (J^P)$ & M (MeV) &    Hadron & $I (J^P)$ & M (MeV)    \\
  \hline    $K^{0}$ & $\frac{1}{2}(0^-)$ & $497.611$  &    $K^{\pm}$ & $\frac{1}{2}(0^-)$ & $493.677$ \\
$\bar{D}^{0}$ & $1/2(0^-)$ & $1864.84$  &    $D^{-}$ & $1/2(0^-)$ & $1869.66$      \\
$\bar{D}^{\ast0}$ & $1/2(1^-)$ & $2006.85$ &  $D^{\ast-}$ & $1/2(1^-)$ & $2010.26$   \\
$D_{1}^{0}$ & $1/2(1^+)$ & $2422.06$ &  $D_{1}^{\pm}$ & $1/2(1^+)$ & $2426.06$  \\
     $D_s^{\pm}$ & $0(0^{-})$ & $1968.35$ &  $D_s^{\ast\pm}$ & $0(1^{-})$ & $2112.20$    \\
$Y(4230)$ & $0(1^-)$ & $4222.70$ &  $Y(4360)$ & $0(1^-)$ & $4372.00$  \\
  $B^{\pm}$ & $\frac{1}{2}(0^-)$ & $5279.41$ &  $B^0$ & $\frac{1}{2}(0^-)$ & $5279.72$  \\
 \hline \hline
\end{tabular}
\label{tab:masses}
\end{table}

\subsection{Branching fractions of $B \to Y(4230)/Y(4360)K$}

\begin{figure}[htb!]
\begin{tabular}{ccc}
  \centering
\includegraphics[width=1.1\textwidth, trim=1cm 11cm 1cm 11cm, clip]{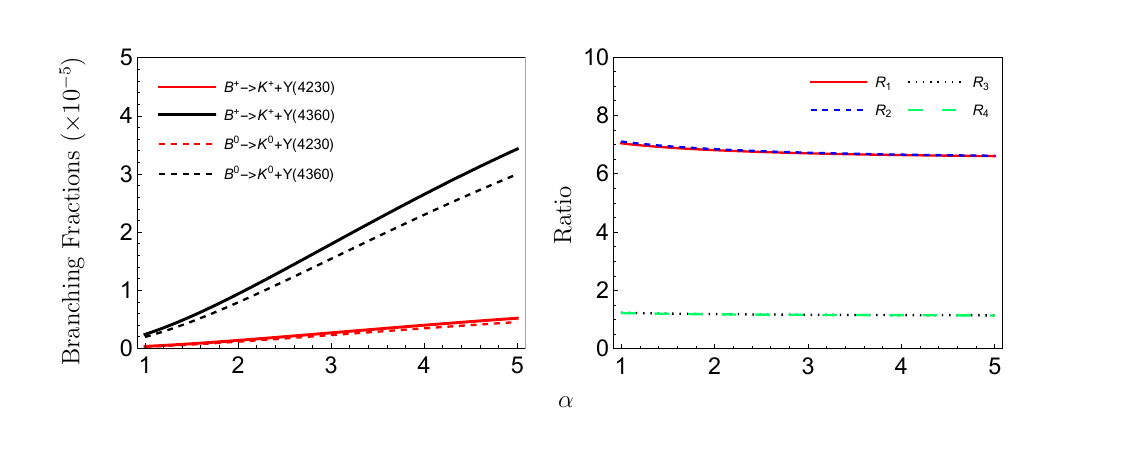}
 \end{tabular}
  \caption{Branching fractions of $B \rightarrow K  Y(4230)$ and $B \rightarrow K  Y(4360)$ (left panel) and their ratios (right panel) as a function of $\alpha$}\label{Fig:br_ratio}
\end{figure}

In Fig.~\ref{Fig:br_ratio}, we present the branching fractions  of the decays  $B \rightarrow K  Y(4230)$ and $B  \rightarrow K  Y(4360)$  as well as  their ratios as the function of  $\alpha$. As shown in the left panel of Fig.~\ref{Fig:br_ratio}, the branching fractions of $\mathcal{B}[B \rightarrow K Y(4230)]$ and  $\mathcal{B}[B \rightarrow K Y(4360)]$ both increase with $\alpha$.  As $\alpha$ is varied from $1$ to $5$, the branching fractions  $\mathcal{B}[B^{+} \rightarrow K^{+} Y(4230)]$ and  $\mathcal{B}[B^{+} \rightarrow K^{+} Y(4360)]$ range from $(0.34 \sim 5.19)\times10^{-6}$  to $(0.24\sim 3.43)\times10^{-5}$, respectively.  For their isospin counterparts,  the branching fractions of  $\mathcal{B}[B^{0} \rightarrow K^{0} Y(4230)]$ and  $\mathcal{B}[B^{0} \rightarrow K^{0} Y(4360)]$ are  $(0.28 \sim 4.53)\times10^{-6}$  and $(0.2\sim 3)\times10^{-5}$, reflecting isospin symmetry. A recent study has shown that $Y(4230)$  and  $\psi(4160)$ are the same $\psi(2~^3D_1)$ state from analysis of the mass distributions~\cite{Zhou:2023yjv}, and therefore the branching fraction $\mathcal{B}[B^+ \to Y(4230)K^+]$ is expected to  be of an order similar to the  branching fraction $\mathcal{B}[B^+ \to \psi(4160)K^+]=(5.1^{+1.3}_{-1.2}\pm 3.0)\times 10^{-4}$, which is larger than that of  $Y(4230)$ as a molecule by two orders of magnitude, implying that $Y(4230)$ and $\psi(4160)$ cannot be assigned as the same state from the perspective of B decays~\cite{LHCb:2013ywr}.

The absolute branching fractions of the above weak decays are dependent on $\alpha$, while their ratios would be weakly dependent on $\alpha$. Next we define following ratios:
\begin{eqnarray}
R_1&=&\frac{\mathcal{B}[B^+ \rightarrow K^+ Y(4360)]}{\mathcal{B}[B^+ \rightarrow K^+ Y(4230)]},  ~~~~
R_2=\frac{\mathcal{B}[B^0 \rightarrow K^0 Y(4360)]}{\mathcal{B}[B^0 \rightarrow K^0 Y(4230)]},\nonumber\\
R_3&=&\frac{\mathcal{B}[B^+ \rightarrow K^+ Y(4230)]}{\mathcal{B}[B^0 \rightarrow K^0 Y(4230)]},\nonumber ~~~~
R_4=\frac{\mathcal{B}[B^+ \rightarrow K^+ Y(4360)]}{\mathcal{B}[B^0 \rightarrow K^0 Y(4360)]}.\nonumber
\end{eqnarray}
In the right panel of Fig.~\ref{Fig:br_ratio}, we show the above ratios as a function of $\alpha$ varying from $1$ to $5$. Our results show  that the ratios  of $R_1$, $R_2$, $R_3$ and $R_4$ are in the range of $7.046\sim6.606$, $7.107\sim6.618$, $1.238\sim1.147$ and $1.227\sim1.145$, respectively. The ratios of $R_1$ and $R_2$ are around $7$, indicating that  the production rates of  $Y(4360)$ in $B$ decays  are  larger than those of  $Y(4230)$ in $B$ decays, which are similar to  the  ratios of  production rates of $D_{s1}(2460)$ to those of  $D_{s0}^*(2317)$ in heavy hadron decays~\cite{Liu:2023cwk}, where $D_{s1}(2460)$ and $D_{s0}^*(2317)$ are also assumed as the HQSS doublet hadronic molecules of $D^*K$ and $DK$.  In our previous studies, the isospin breaking effect of the production rates of $X(3872)$ in $B^+$ and $B^0$ decays was shown to be more than $50\%$, which is mainly attributed to the isospin breaking of  $X(3872)$ coupling to the neutral channel and electron channel~\cite{Wu:2023rrp}.  However, since the isospin breaking of  $Y(4230)$ and $Y(4360)$ couplings to the neutral channels and electric channels in Fig.~\ref{Fig:2} is small due to the deeply bound energies, the ratios of $R_3$ and $R_4$ are estimated to be around $1$, indicating the small isospin breaking effect of the production of $Y(4230)$ and $Y(4360)$  in $B^+$ and $B^0$ decays.

From our results, we can see that the branching fractions of $\mathcal{B}[B \rightarrow K  Y(4230)]$ and $\mathcal{B}[B  \rightarrow K  Y(4360)]$ are up the order of  $10^{-6}$ and  $10^{-5}$, respectively.  In Ref.~\cite{Gao:2017sqa}, the branching fraction of the decay $Y(4230)\to J/\psi \pi^+\pi^-$ was estimated to be around $10\%$, and the branching fraction $\mathcal{B}[B \rightarrow  ( Y(4230)\to J/\psi \pi^+\pi^- )K]$ was estimated at around $10^{-7}$. Taking into account the branching fraction $\mathcal{B}(B \rightarrow  J/\psi \pi^+\pi^- K)\sim 10^{-3}$, one obtains a particularly low ratio  $\mathcal{B}[B \rightarrow  ( Y(4230)\to J/\psi \pi^+\pi^- )K]/\mathcal{B}(B \rightarrow  J/\psi \pi^+\pi^- K)\sim 10^{-4}$, which implies that larger statistics are needed to observe the significant signal of  $Y(4230)$ in $B$ decays.  Similarly, the branching fraction of the decay $Y(4360)\to J/\psi \pi^+\pi^-$ is estimated to be around $10\%$~\cite{Zhang:2018zog}, and  we can obtain the  ratio $\mathcal{B}[B \rightarrow  ( Y(4360)\to J/\psi \pi^+\pi^- )K]/\mathcal{B}(B \rightarrow  J/\psi \pi^+\pi^- K)\sim 10^{-3}$, larger than that of $Y(4230)$ by one order of magnitude, which indicates that $Y(4360)$ is more likely to be detected than $Y(4230)$ in the decay $B \rightarrow  J/\psi \pi^+\pi^- K$.

\subsection{Decay constants of $Y(4230)/Y(4360)$}

\begin{figure}[ttt]
\begin{tabular}{c}
  \centering
 \includegraphics[width=11.0cm]{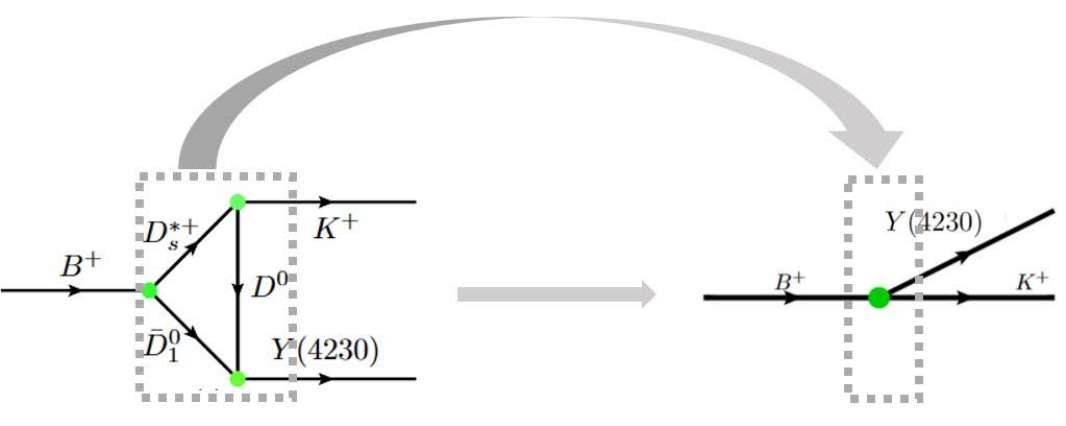}
 \end{tabular}
  \caption{Triangle diagrams illustrating  the decays $B \to K Y(4230)$ simplified as tree diagrams}\label{Fig:TritoTree}
\end{figure}

In this section, we discuss the decay constants of $Y(4230)$ and $Y(4360)$ extracted from the $B$ decays and the electronic decays.  In the former case,  the decay constant can be extracted from the equivalence between the triangle diagram and the tree diagram, as illustrated in  Fig.~\ref{Fig:TritoTree}, which was applied to extract the decay constant of $X(3872)$~\cite{Wu:2023rrp}.

With the factorization ansatz, the weak decay amplitudes of $B(k_0)\to K(p_1)Y(p_2)$ actually can be expressed as the product of two matrix elements:
\begin{eqnarray}\label{Ds-KK1}
\mathcal{A}\left(B\to Y K\right)&=&\frac{G_{F}}{\sqrt{2}} V_{cb}V_{cs} a_{2}\left\langle Y|J^{\mu}| 0\right\rangle\left\langle K|J_{\mu}| B\right\rangle , \label{Eq:BYK}
\end{eqnarray}
where the matrix elements are
\begin{eqnarray}
\left\langle Y|J^{\mu}| 0\right\rangle =\varepsilon^{\mu} f_{Y} m_{Y},   \label{Eq:0Y}
\end{eqnarray}
and
\begin{eqnarray}
\left\langle K| J_{\mu}| B\right\rangle =\left[(p_{B}+p_{K})_{\mu}-\frac{m_{B}^2-m_{K}^2}{q^2}q_{\mu}\right] F_{1}(q^2)+\frac{m_{B}^2-m_{K}^2}{q^2}q_{\mu} F_{0}(q^2),\label{Eq:BK}
\end{eqnarray}
with $F_0$ and $F_1$ being  the form factors for $B(k_0)\to K(q_2)$,  and  $P=k_0+q_2$ and $q=k_0-q_2$. The form factors  $F_0$ and $F_1$ are parameterized by Eq.~(\ref{Eq:A1}), and the corresponding  values of $F(0)$, $a$, and $b$ in the form factors  are collected in Table~\ref{BtoDformfactor}.

\begin{figure}[htb!]
\begin{tabular}{ccc}
  \centering
 \includegraphics[width=1.1\textwidth, trim=1.1cm 10.5cm 0.2cm 10cm, clip]{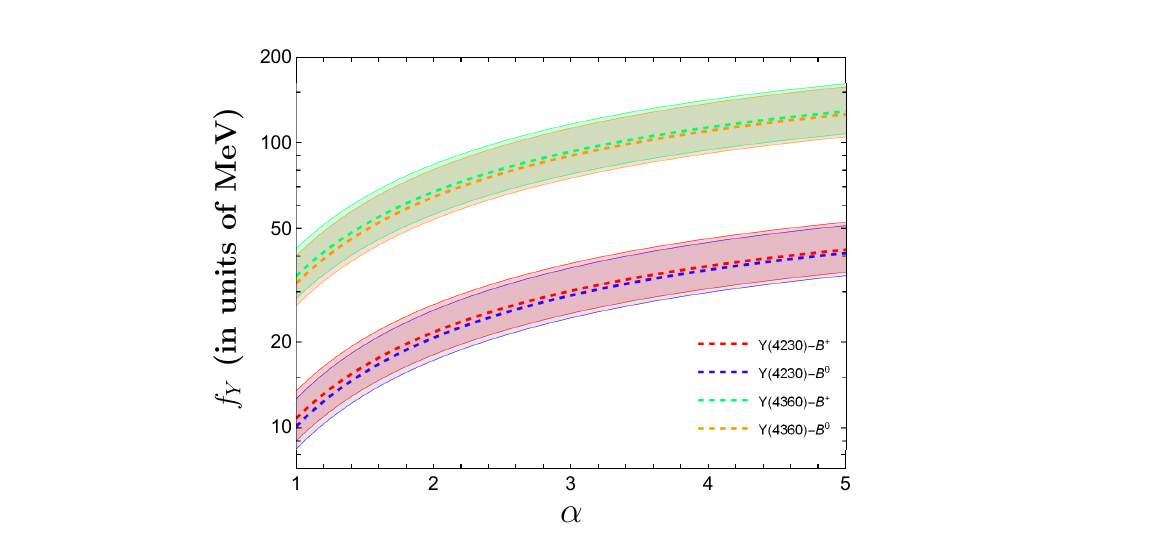}
 \end{tabular}
  \caption{  Decay constants of $Y(4230)$ and $Y(4360)$ extracted in $B$ decays as a function of $\alpha$ (in units of MeV). The dashed lines and bands represent their central values and corresponding uncertainties}\label{Fig:fY}
\end{figure}

With Eqs.~(\ref{Eq:BYK})-(\ref{Eq:BK}), the weak decay amplitude of $B(k_0) \to Y(q_1) K(q_2)$ has the form

\begin{eqnarray}\label{BtoX387245}
\mathcal{A}(B \to Y K)=\frac{G_{F}}{\sqrt{2}} V_{cb}V_{cs} a_{2} m_Y f_Y (k_0+q_2)\cdot \epsilon(q_1) F_1(q^2_1),
\end{eqnarray}
where the effective Wilson coefficient $a_2=0.271$ is determined by reproducing the branching fraction of the decay $B\to J/\psi K$\footnote{ From the decays of   $B\to J/\psi K^*$, $B_s\to J/\psi \phi $,  $B_s \to J/\psi \eta$, and $B \to \psi(2S) K$~\cite{Wu:2023rrp}, the  extracted value of the effective Wilson coefficient $a_2$ is around $0.271$. Therefore, we assume that the effective Wilson coefficient of a $B$ meson decaying into vector charmonium/charmonium-like states shares a similar value.   }.  From the derived value of $a_2$, we set a $20\%$ uncertainty in the following calculations. Based on the equivalence between the triangle diagram and the tree diagram,  the decay constants of $Y(4230)$ and $Y(4360)$ as a function of $\alpha$ are shown in Fig.~\ref{Fig:fY}. As $\alpha$ varies from $1$ to $5$, the decay constants of $Y(4230)$ and $Y(4360)$ extracted in $B^+$ decays are in the ranges of  $10.8^{+2.69}_{-1.79}\sim 42.1^{+10.5}_{-7.0}$  MeV and  $34.0^{+8.47}_{-5.66}\sim 128.6^{+32.0}_{-21.4}$  MeV. According to  isospin symmetry, the  decay constants of $Y(4230)$ and $Y(4360)$ extracted  in  $B^0$ decays are similar to those in  $B^+$ decays. Our results indicate that the decay constant of $Y(4360)$ is larger than that of $Y(4230)$, similar to the decay constants of $D_{s1}(2460)$ and $D_{s0}^*(2317)$~\cite{Liu:2023cwk}.

On the other hand, the decay constant can also be extracted from the electronic widths of the vector charmonium states via the decay $Y\to\gamma^*\to e^+e^-$.  The vector charmonium state coupling to the photon is described by vector meson dominance model, and the corresponding  Lagrangian  is written as~\cite{Korchin:2011ze,Shi:2023ntq}
\begin{widetext}
\begin{eqnarray}\label{Ygamma}
\mathcal{L}_{Y\gamma}=-\frac{e}{2}\frac{f_Y Q_c}{m_Y}F^{\mu\nu}Y_{\mu\nu},
\end{eqnarray}
\end{widetext}
with electric charge of the charm quark being  $Q_c=2/3$, where $f_Y$ represents the  decay constant of the $Y$ state.  The width of a vector charmonium state decaying into a pair of electron and positron is derived  as
\begin{eqnarray}\label{gamma_ee}
\Gamma^{Y \to e^+e^-}=\frac{4\pi \alpha}{3}\frac{f_Y^2 Q_c^2}{m_Y},
\end{eqnarray}
\label{electroc}
where $\alpha$ is the fine-structure constant. If one can obtain the electronic decay width of a $Y$ state, its decay constant can be derived from Eq.~(\ref{gamma_ee}).

\begin{table}[!h]
\centering
\caption{ Decay constants of $Y(4230)$ and $Y(4360)$ obtained  via the $B$ decays in present work and the electronic widths in Ref.~\cite{Zhang:2018zog} (in units of MeV), where the error in the former scenario  come from the uncertainty of the effective Wilson coefficient $a_2$ \label{resultsbsto2317}
}
\begin{tabular}{c c c  c c c c c c c}
  \hline \hline
State      &~~~~
Decay constant(B decay)   &~~~~
Decay constant(decay to $e^+e^-$)~\cite{Zhang:2018zog}
         \\ \hline
      $Y(4230)$      & $10.8^{+2.69}_{-1.79}\sim 42.1^{+10.5}_{-7.0}$    & $39.4\sim 93.8$
         \\
      $Y(4360)$     & $34.0^{+8.47}_{-5.66}\sim 128.6^{+32.0}_{-21.4}$   & $73.9\sim 210.2$
         \\
  \hline \hline
\end{tabular}
\end{table}

At present, there exists no experimental data for the electronic  decay widths of  $Y(4230)\to e^+e^-$  and  $Y(4360)\to e^+e^-$. In Ref.~\cite{Zhang:2018zog}, a combined fit was performed on the cross section of $e^+e^-\to \omega\chi_{c0}$, $\pi^+\pi^- h_c$, $\pi^+\pi^- J/\psi$, $\pi^+\pi^- \psi(3686)$ and $\pi^+ D^0 D^{*-}+c.c.$, obtaining the electronic  decay widths for $Y(4230)\to e^+e^-$  and  $Y(4360)\to e^+e^-$ of $36.4\sim206.6$ eV and $123.8\sim1001.7$ eV, respectively, consistent with Ref.~\cite{Gao:2017sqa} and within the upper limit of $Y(4230)$ in Ref.~\cite{Qin:2016spb}. Identifying the electronic decay widths in Ref.~\cite{Zhang:2018zog} as inputs,  we determine the  decay constants of $Y(4230)$ and $Y(4390)$ as shown in Table~\ref{resultsbsto2317}.   We can see that all of the decay constants  of  $Y(4230)$ and $Y(4360)$ obtained in $B$ decays are smaller than those obtained in  electronic decays.    It  should be noted that  a crude assumption, in which the non-factorization contribution of  the $B$ meson decaying into charmonium/charmonium-like states and kaon meson is absorbed into the effective Wilson coefficient $a_2$, may bring the uncertainties for our results.

\section{Summary}
\label{sec:summary}

Two charmonium-like states $Y(4230)$ and $Y(4360)$ splitting from the previously discovered $Y(4260)$ imply a fine structure in this energy sector, which can be well explained by the HQSS doublet hadronic molecules of $D\bar{D}_1$ and $D^*\bar{D}_1$. Many conventional charmonium states, such as $\psi(3770)$, $\psi(4040)$,  $\psi(4160)$, and $\psi(4415)$, are observed in $B$ decays, while $Y(4230)$ and $Y(4360)$ are absent. In this work, assuming $Y(4230)$ and $Y(4360)$ as the bound states of  $D\bar{D}_1$ and $D^*\bar{D}_1$, we employ the triangle diagram mechanism to illustrate the production of $Y(4230)$ and $Y(4360)$ in $B$ decays, and then employ the effective Lagrangian approach to calculate the branching fractions of the decays  $B \rightarrow K Y(4230)$ and $B \rightarrow K Y(4360)$. Moreover, according to the equivalence between the triangle diagram and the tree diagram, we extract the decay constants of $Y(4230)$ and $Y(4360)$ in $B$ decays.

Our results show that the branching fractions $\mathcal{B}[B \rightarrow K Y(4230)]$ and  $\mathcal{B}[B \rightarrow K Y(4360)]$ are up to the order of $10^{-6}$ and $10^{-5}$, respectively, smaller than those of $\mathcal{B}[B \rightarrow K \psi(4160)]$ and $\mathcal{B}[B \rightarrow K \psi(4415)]$ by at least one order of magnitude, which might explain why $Y(4230)$ and $Y(4360)$ are not observed in $B$ decays. Designating the branching fraction of the decays  $Y(4230) \to J/\psi \pi^+\pi^-$ and $Y(4360)\to J/\psi \pi^+\pi^-$ as $10\%$, the ratios of branching fractions $\mathcal{B}[B \rightarrow  Y(4230) K \to  J/\psi \pi^+\pi^-K]$ and  $\mathcal{B}[B \rightarrow K Y(4360)\to  J/\psi \pi^+\pi^-K]$ to $\mathcal{B}(B \rightarrow  J/\psi \pi^+\pi^- K)$ are up to an order of $10^{-4}$ and $10^{-3}$, respectively, which indicates that larger statistics are needed to observe the  $Y(4230)$ and $Y(4360)$ in  $B \rightarrow  J/\psi \pi^+\pi^- K$ decay. Finally, the   decay constants of $Y(4230)$ and $Y(4360)$ extracted  in $B$ decays are in the ranges of $10.8^{+2.69}_{-1.79}\sim 42.1^{+10.5}_{-7.0}$~MeV and $34.0^{+8.47}_{-5.66}\sim 128.6^{+32.0}_{-21.4}$~MeV, respectively,  a bit smaller than those extracted in electronic decays.

\section{Acknowledgments}

This work is supported by the National Natural Science Foundation of China under Grants No.12105007 and 12405093.

\bibliography{biblio}

%merlin.mbs apsrev4-1.bst 2010-07-25 4.21a (PWD, AO, DPC) hacked
%Control: key (0)
%Control: author (8) initials jnrlst
%Control: editor formatted (1) identically to author
%Control: production of article title (-1) disabled
%Control: page (0) single
%Control: year (1) truncated
%Control: production of eprint (0) enabled
\begin{thebibliography}{92}%
\makeatletter
\providecommand \@ifxundefined [1]{%
 \@ifx{#1\undefined}
}%
\providecommand \@ifnum [1]{%
 \ifnum #1\expandafter \@firstoftwo
 \else \expandafter \@secondoftwo
 \fi
}%
\providecommand \@ifx [1]{%
 \ifx #1\expandafter \@firstoftwo
 \else \expandafter \@secondoftwo
 \fi
}%
\providecommand \natexlab [1]{#1}%
\providecommand \enquote  [1]{``#1''}%
\providecommand \bibnamefont  [1]{#1}%
\providecommand \bibfnamefont [1]{#1}%
\providecommand \citenamefont [1]{#1}%
\providecommand \href@noop [0]{\@secondoftwo}%
\providecommand \href [0]{\begingroup \@sanitize@url \@href}%
\providecommand \@href[1]{\@@startlink{#1}\@@href}%
\providecommand \@@href[1]{\endgroup#1\@@endlink}%
\providecommand \@sanitize@url [0]{\catcode `\\12\catcode `\$12\catcode
  `\&12\catcode `\#12\catcode `\^12\catcode `\_12\catcode `\%12\relax}%
\providecommand \@@startlink[1]{}%
\providecommand \@@endlink[0]{}%
\providecommand \url  [0]{\begingroup\@sanitize@url \@url }%
\providecommand \@url [1]{\endgroup\@href {#1}{\urlprefix }}%
\providecommand \urlprefix  [0]{URL }%
\providecommand \Eprint [0]{\href }%
\providecommand \doibase [0]{http://dx.doi.org/}%
\providecommand \selectlanguage [0]{\@gobble}%
\providecommand \bibinfo  [0]{\@secondoftwo}%
\providecommand \bibfield  [0]{\@secondoftwo}%
\providecommand \translation [1]{[#1]}%
\providecommand \BibitemOpen [0]{}%
\providecommand \bibitemStop [0]{}%
\providecommand \bibitemNoStop [0]{.\EOS\space}%
\providecommand \EOS [0]{\spacefactor3000\relax}%
\providecommand \BibitemShut  [1]{\csname bibitem#1\endcsname}%
\let\auto@bib@innerbib\@empty
%</preamble>
\bibitem [{\citenamefont {Caswell}\ and\ \citenamefont
  {Lepage}(1986)}]{Caswell:1985ui}%
  \BibitemOpen
  \bibfield  {author} {\bibinfo {author} {\bibfnamefont {W.~E.}\ \bibnamefont
  {Caswell}}\ and\ \bibinfo {author} {\bibfnamefont {G.~P.}\ \bibnamefont
  {Lepage}},\ }\href {\doibase 10.1016/0370-2693(86)91297-9} {\bibfield
  {journal} {\bibinfo  {journal} {Phys. Lett. B}\ }\textbf {\bibinfo {volume}
  {167}},\ \bibinfo {pages} {437} (\bibinfo {year} {1986})}\BibitemShut
  {NoStop}%
\bibitem [{\citenamefont {Pineda}\ and\ \citenamefont
  {Soto}(1998)}]{Pineda:1997bj}%
  \BibitemOpen
  \bibfield  {author} {\bibinfo {author} {\bibfnamefont {A.}~\bibnamefont
  {Pineda}}\ and\ \bibinfo {author} {\bibfnamefont {J.}~\bibnamefont {Soto}},\
  }\href {\doibase 10.1016/S0920-5632(97)01102-X} {\bibfield  {journal}
  {\bibinfo  {journal} {Nucl. Phys. B Proc. Suppl.}\ }\textbf {\bibinfo
  {volume} {64}},\ \bibinfo {pages} {428} (\bibinfo {year} {1998})},\ \Eprint
  {http://arxiv.org/abs/hep-ph/9707481} {arXiv:hep-ph/9707481} \BibitemShut
  {NoStop}%
\bibitem [{\citenamefont {Eichten}\ \emph {et~al.}(1975)\citenamefont
  {Eichten}, \citenamefont {Gottfried}, \citenamefont {Kinoshita},
  \citenamefont {Kogut}, \citenamefont {Lane},\ and\ \citenamefont
  {Yan}}]{Eichten:1974af}%
  \BibitemOpen
  \bibfield  {author} {\bibinfo {author} {\bibfnamefont {E.}~\bibnamefont
  {Eichten}}, \bibinfo {author} {\bibfnamefont {K.}~\bibnamefont {Gottfried}},
  \bibinfo {author} {\bibfnamefont {T.}~\bibnamefont {Kinoshita}}, \bibinfo
  {author} {\bibfnamefont {J.~B.}\ \bibnamefont {Kogut}}, \bibinfo {author}
  {\bibfnamefont {K.~D.}\ \bibnamefont {Lane}}, \ and\ \bibinfo {author}
  {\bibfnamefont {T.-M.}\ \bibnamefont {Yan}},\ }\href {\doibase
  10.1103/PhysRevLett.34.369} {\bibfield  {journal} {\bibinfo  {journal} {Phys.
  Rev. Lett.}\ }\textbf {\bibinfo {volume} {34}},\ \bibinfo {pages} {369}
  (\bibinfo {year} {1975})},\ \bibinfo {note} {[Erratum: Phys.Rev.Lett. 36,
  1276 (1976)]}\BibitemShut {NoStop}%
\bibitem [{\citenamefont {Godfrey}\ and\ \citenamefont
  {Isgur}(1985)}]{Godfrey:1985xj}%
  \BibitemOpen
  \bibfield  {author} {\bibinfo {author} {\bibfnamefont {S.}~\bibnamefont
  {Godfrey}}\ and\ \bibinfo {author} {\bibfnamefont {N.}~\bibnamefont
  {Isgur}},\ }\href {\doibase 10.1103/PhysRevD.32.189} {\bibfield  {journal}
  {\bibinfo  {journal} {Phys. Rev. D}\ }\textbf {\bibinfo {volume} {32}},\
  \bibinfo {pages} {189} (\bibinfo {year} {1985})}\BibitemShut {NoStop}%
\bibitem [{\citenamefont {Li}\ and\ \citenamefont {Chao}(2009)}]{Li:2009zu}%
  \BibitemOpen
  \bibfield  {author} {\bibinfo {author} {\bibfnamefont {B.-Q.}\ \bibnamefont
  {Li}}\ and\ \bibinfo {author} {\bibfnamefont {K.-T.}\ \bibnamefont {Chao}},\
  }\href {\doibase 10.1103/PhysRevD.79.094004} {\bibfield  {journal} {\bibinfo
  {journal} {Phys. Rev. D}\ }\textbf {\bibinfo {volume} {79}},\ \bibinfo
  {pages} {094004} (\bibinfo {year} {2009})},\ \Eprint
  {http://arxiv.org/abs/0903.5506} {arXiv:0903.5506 [hep-ph]} \BibitemShut
  {NoStop}%
\bibitem [{\citenamefont {Ortega}\ \emph {et~al.}(2010)\citenamefont {Ortega},
  \citenamefont {Segovia}, \citenamefont {Entem},\ and\ \citenamefont
  {Fernandez}}]{Ortega:2009hj}%
  \BibitemOpen
  \bibfield  {author} {\bibinfo {author} {\bibfnamefont {P.~G.}\ \bibnamefont
  {Ortega}}, \bibinfo {author} {\bibfnamefont {J.}~\bibnamefont {Segovia}},
  \bibinfo {author} {\bibfnamefont {D.~R.}\ \bibnamefont {Entem}}, \ and\
  \bibinfo {author} {\bibfnamefont {F.}~\bibnamefont {Fernandez}},\ }\href
  {\doibase 10.1103/PhysRevD.81.054023} {\bibfield  {journal} {\bibinfo
  {journal} {Phys. Rev. D}\ }\textbf {\bibinfo {volume} {81}},\ \bibinfo
  {pages} {054023} (\bibinfo {year} {2010})},\ \Eprint
  {http://arxiv.org/abs/0907.3997} {arXiv:0907.3997 [hep-ph]} \BibitemShut
  {NoStop}%
\bibitem [{\citenamefont {Tan}\ and\ \citenamefont {Ping}(2019)}]{Tan:2019qwe}%
  \BibitemOpen
  \bibfield  {author} {\bibinfo {author} {\bibfnamefont {Y.}~\bibnamefont
  {Tan}}\ and\ \bibinfo {author} {\bibfnamefont {J.}~\bibnamefont {Ping}},\
  }\href {\doibase 10.1103/PhysRevD.100.034022} {\bibfield  {journal} {\bibinfo
   {journal} {Phys. Rev. D}\ }\textbf {\bibinfo {volume} {100}},\ \bibinfo
  {pages} {034022} (\bibinfo {year} {2019})},\ \Eprint
  {http://arxiv.org/abs/1906.09690} {arXiv:1906.09690 [hep-ph]} \BibitemShut
  {NoStop}%
\bibitem [{\citenamefont {Duan}\ \emph {et~al.}(2020)\citenamefont {Duan},
  \citenamefont {Luo}, \citenamefont {Liu},\ and\ \citenamefont
  {Matsuki}}]{Duan:2020tsx}%
  \BibitemOpen
  \bibfield  {author} {\bibinfo {author} {\bibfnamefont {M.-X.}\ \bibnamefont
  {Duan}}, \bibinfo {author} {\bibfnamefont {S.-Q.}\ \bibnamefont {Luo}},
  \bibinfo {author} {\bibfnamefont {X.}~\bibnamefont {Liu}}, \ and\ \bibinfo
  {author} {\bibfnamefont {T.}~\bibnamefont {Matsuki}},\ }\href {\doibase
  10.1103/PhysRevD.101.054029} {\bibfield  {journal} {\bibinfo  {journal}
  {Phys. Rev. D}\ }\textbf {\bibinfo {volume} {101}},\ \bibinfo {pages}
  {054029} (\bibinfo {year} {2020})},\ \Eprint
  {http://arxiv.org/abs/2002.03311} {arXiv:2002.03311 [hep-ph]} \BibitemShut
  {NoStop}%
\bibitem [{\citenamefont {Deng}\ \emph {et~al.}(2024)\citenamefont {Deng},
  \citenamefont {Ni}, \citenamefont {Li},\ and\ \citenamefont
  {Zhong}}]{Deng:2023mza}%
  \BibitemOpen
  \bibfield  {author} {\bibinfo {author} {\bibfnamefont {Q.}~\bibnamefont
  {Deng}}, \bibinfo {author} {\bibfnamefont {R.-H.}\ \bibnamefont {Ni}},
  \bibinfo {author} {\bibfnamefont {Q.}~\bibnamefont {Li}}, \ and\ \bibinfo
  {author} {\bibfnamefont {X.-H.}\ \bibnamefont {Zhong}},\ }\href {\doibase
  10.1103/PhysRevD.110.056034} {\bibfield  {journal} {\bibinfo  {journal}
  {Phys. Rev. D}\ }\textbf {\bibinfo {volume} {110}},\ \bibinfo {pages}
  {056034} (\bibinfo {year} {2024})},\ \Eprint
  {http://arxiv.org/abs/2312.10296} {arXiv:2312.10296 [hep-ph]} \BibitemShut
  {NoStop}%
\bibitem [{\citenamefont {Bali}\ \emph {et~al.}(2005)\citenamefont {Bali},
  \citenamefont {Neff}, \citenamefont {Duessel}, \citenamefont {Lippert},\ and\
  \citenamefont {Schilling}}]{Bali:2005fu}%
  \BibitemOpen
  \bibfield  {author} {\bibinfo {author} {\bibfnamefont {G.~S.}\ \bibnamefont
  {Bali}}, \bibinfo {author} {\bibfnamefont {H.}~\bibnamefont {Neff}}, \bibinfo
  {author} {\bibfnamefont {T.}~\bibnamefont {Duessel}}, \bibinfo {author}
  {\bibfnamefont {T.}~\bibnamefont {Lippert}}, \ and\ \bibinfo {author}
  {\bibfnamefont {K.}~\bibnamefont {Schilling}} (\bibinfo {collaboration}
  {SESAM}),\ }\href {\doibase 10.1103/PhysRevD.71.114513} {\bibfield  {journal}
  {\bibinfo  {journal} {Phys. Rev. D}\ }\textbf {\bibinfo {volume} {71}},\
  \bibinfo {pages} {114513} (\bibinfo {year} {2005})},\ \Eprint
  {http://arxiv.org/abs/hep-lat/0505012} {arXiv:hep-lat/0505012} \BibitemShut
  {NoStop}%
\bibitem [{\citenamefont {Barnes}\ \emph {et~al.}(2005)\citenamefont {Barnes},
  \citenamefont {Godfrey},\ and\ \citenamefont {Swanson}}]{Barnes:2005pb}%
  \BibitemOpen
  \bibfield  {author} {\bibinfo {author} {\bibfnamefont {T.}~\bibnamefont
  {Barnes}}, \bibinfo {author} {\bibfnamefont {S.}~\bibnamefont {Godfrey}}, \
  and\ \bibinfo {author} {\bibfnamefont {E.~S.}\ \bibnamefont {Swanson}},\
  }\href {\doibase 10.1103/PhysRevD.72.054026} {\bibfield  {journal} {\bibinfo
  {journal} {Phys. Rev. D}\ }\textbf {\bibinfo {volume} {72}},\ \bibinfo
  {pages} {054026} (\bibinfo {year} {2005})},\ \Eprint
  {http://arxiv.org/abs/hep-ph/0505002} {arXiv:hep-ph/0505002} \BibitemShut
  {NoStop}%
\bibitem [{\citenamefont {Chen}\ \emph
  {et~al.}(2016{\natexlab{a}})\citenamefont {Chen}, \citenamefont {Chen},
  \citenamefont {Liu},\ and\ \citenamefont {Zhu}}]{Chen:2016qju}%
  \BibitemOpen
  \bibfield  {author} {\bibinfo {author} {\bibfnamefont {H.-X.}\ \bibnamefont
  {Chen}}, \bibinfo {author} {\bibfnamefont {W.}~\bibnamefont {Chen}}, \bibinfo
  {author} {\bibfnamefont {X.}~\bibnamefont {Liu}}, \ and\ \bibinfo {author}
  {\bibfnamefont {S.-L.}\ \bibnamefont {Zhu}},\ }\href {\doibase
  10.1016/j.physrep.2016.05.004} {\bibfield  {journal} {\bibinfo  {journal}
  {Phys. Rept.}\ }\textbf {\bibinfo {volume} {639}},\ \bibinfo {pages} {1}
  (\bibinfo {year} {2016}{\natexlab{a}})},\ \Eprint
  {http://arxiv.org/abs/1601.02092} {arXiv:1601.02092 [hep-ph]} \BibitemShut
  {NoStop}%
\bibitem [{\citenamefont {Lebed}\ \emph {et~al.}(2017)\citenamefont {Lebed},
  \citenamefont {Mitchell},\ and\ \citenamefont {Swanson}}]{Lebed:2016hpi}%
  \BibitemOpen
  \bibfield  {author} {\bibinfo {author} {\bibfnamefont {R.~F.}\ \bibnamefont
  {Lebed}}, \bibinfo {author} {\bibfnamefont {R.~E.}\ \bibnamefont {Mitchell}},
  \ and\ \bibinfo {author} {\bibfnamefont {E.~S.}\ \bibnamefont {Swanson}},\
  }\href {\doibase 10.1016/j.ppnp.2016.11.003} {\bibfield  {journal} {\bibinfo
  {journal} {Prog. Part. Nucl. Phys.}\ }\textbf {\bibinfo {volume} {93}},\
  \bibinfo {pages} {143} (\bibinfo {year} {2017})},\ \Eprint
  {http://arxiv.org/abs/1610.04528} {arXiv:1610.04528 [hep-ph]} \BibitemShut
  {NoStop}%
\bibitem [{\citenamefont {Oset}\ \emph {et~al.}(2016)\citenamefont {Oset} \emph
  {et~al.}}]{Oset:2016lyh}%
  \BibitemOpen
  \bibfield  {author} {\bibinfo {author} {\bibfnamefont {E.}~\bibnamefont
  {Oset}} \emph {et~al.},\ }\href {\doibase 10.1142/S0218301316300010}
  {\bibfield  {journal} {\bibinfo  {journal} {Int. J. Mod. Phys. E}\ }\textbf
  {\bibinfo {volume} {25}},\ \bibinfo {pages} {1630001} (\bibinfo {year}
  {2016})},\ \Eprint {http://arxiv.org/abs/1601.03972} {arXiv:1601.03972
  [hep-ph]} \BibitemShut {NoStop}%
\bibitem [{\citenamefont {Esposito}\ \emph {et~al.}(2017)\citenamefont
  {Esposito}, \citenamefont {Pilloni},\ and\ \citenamefont
  {Polosa}}]{Esposito:2016noz}%
  \BibitemOpen
  \bibfield  {author} {\bibinfo {author} {\bibfnamefont {A.}~\bibnamefont
  {Esposito}}, \bibinfo {author} {\bibfnamefont {A.}~\bibnamefont {Pilloni}}, \
  and\ \bibinfo {author} {\bibfnamefont {A.~D.}\ \bibnamefont {Polosa}},\
  }\href {\doibase 10.1016/j.physrep.2016.11.002} {\bibfield  {journal}
  {\bibinfo  {journal} {Phys. Rept.}\ }\textbf {\bibinfo {volume} {668}},\
  \bibinfo {pages} {1} (\bibinfo {year} {2017})},\ \Eprint
  {http://arxiv.org/abs/1611.07920} {arXiv:1611.07920 [hep-ph]} \BibitemShut
  {NoStop}%
\bibitem [{\citenamefont {Dong}\ \emph {et~al.}(2017)\citenamefont {Dong},
  \citenamefont {Faessler},\ and\ \citenamefont {Lyubovitskij}}]{Dong:2017gaw}%
  \BibitemOpen
  \bibfield  {author} {\bibinfo {author} {\bibfnamefont {Y.}~\bibnamefont
  {Dong}}, \bibinfo {author} {\bibfnamefont {A.}~\bibnamefont {Faessler}}, \
  and\ \bibinfo {author} {\bibfnamefont {V.~E.}\ \bibnamefont {Lyubovitskij}},\
  }\href {\doibase 10.1016/j.ppnp.2017.01.002} {\bibfield  {journal} {\bibinfo
  {journal} {Prog. Part. Nucl. Phys.}\ }\textbf {\bibinfo {volume} {94}},\
  \bibinfo {pages} {282} (\bibinfo {year} {2017})}\BibitemShut {NoStop}%
\bibitem [{\citenamefont {Guo}\ \emph {et~al.}(2018)\citenamefont {Guo},
  \citenamefont {Hanhart}, \citenamefont {Mei\ss{}ner}, \citenamefont {Wang},
  \citenamefont {Zhao},\ and\ \citenamefont {Zou}}]{Guo:2017jvc}%
  \BibitemOpen
  \bibfield  {author} {\bibinfo {author} {\bibfnamefont {F.-K.}\ \bibnamefont
  {Guo}}, \bibinfo {author} {\bibfnamefont {C.}~\bibnamefont {Hanhart}},
  \bibinfo {author} {\bibfnamefont {U.-G.}\ \bibnamefont {Mei\ss{}ner}},
  \bibinfo {author} {\bibfnamefont {Q.}~\bibnamefont {Wang}}, \bibinfo {author}
  {\bibfnamefont {Q.}~\bibnamefont {Zhao}}, \ and\ \bibinfo {author}
  {\bibfnamefont {B.-S.}\ \bibnamefont {Zou}},\ }\href {\doibase
  10.1103/RevModPhys.90.015004} {\bibfield  {journal} {\bibinfo  {journal}
  {Rev. Mod. Phys.}\ }\textbf {\bibinfo {volume} {90}},\ \bibinfo {pages}
  {015004} (\bibinfo {year} {2018})},\ \bibinfo {note} {[Erratum: Rev.Mod.Phys.
  94, 029901 (2022)]},\ \Eprint {http://arxiv.org/abs/1705.00141}
  {arXiv:1705.00141 [hep-ph]} \BibitemShut {NoStop}%
\bibitem [{\citenamefont {Olsen}\ \emph {et~al.}(2018)\citenamefont {Olsen},
  \citenamefont {Skwarnicki},\ and\ \citenamefont {Zieminska}}]{Olsen:2017bmm}%
  \BibitemOpen
  \bibfield  {author} {\bibinfo {author} {\bibfnamefont {S.~L.}\ \bibnamefont
  {Olsen}}, \bibinfo {author} {\bibfnamefont {T.}~\bibnamefont {Skwarnicki}}, \
  and\ \bibinfo {author} {\bibfnamefont {D.}~\bibnamefont {Zieminska}},\ }\href
  {\doibase 10.1103/RevModPhys.90.015003} {\bibfield  {journal} {\bibinfo
  {journal} {Rev. Mod. Phys.}\ }\textbf {\bibinfo {volume} {90}},\ \bibinfo
  {pages} {015003} (\bibinfo {year} {2018})},\ \Eprint
  {http://arxiv.org/abs/1708.04012} {arXiv:1708.04012 [hep-ph]} \BibitemShut
  {NoStop}%
\bibitem [{\citenamefont {Karliner}\ \emph {et~al.}(2018)\citenamefont
  {Karliner}, \citenamefont {Rosner},\ and\ \citenamefont
  {Skwarnicki}}]{Karliner:2017qhf}%
  \BibitemOpen
  \bibfield  {author} {\bibinfo {author} {\bibfnamefont {M.}~\bibnamefont
  {Karliner}}, \bibinfo {author} {\bibfnamefont {J.~L.}\ \bibnamefont
  {Rosner}}, \ and\ \bibinfo {author} {\bibfnamefont {T.}~\bibnamefont
  {Skwarnicki}},\ }\href {\doibase 10.1146/annurev-nucl-101917-020902}
  {\bibfield  {journal} {\bibinfo  {journal} {Ann. Rev. Nucl. Part. Sci.}\
  }\textbf {\bibinfo {volume} {68}},\ \bibinfo {pages} {17} (\bibinfo {year}
  {2018})},\ \Eprint {http://arxiv.org/abs/1711.10626} {arXiv:1711.10626
  [hep-ph]} \BibitemShut {NoStop}%
\bibitem [{\citenamefont {Brambilla}\ \emph {et~al.}(2020)\citenamefont
  {Brambilla}, \citenamefont {Eidelman}, \citenamefont {Hanhart}, \citenamefont
  {Nefediev}, \citenamefont {Shen}, \citenamefont {Thomas}, \citenamefont
  {Vairo},\ and\ \citenamefont {Yuan}}]{Brambilla:2019esw}%
  \BibitemOpen
  \bibfield  {author} {\bibinfo {author} {\bibfnamefont {N.}~\bibnamefont
  {Brambilla}}, \bibinfo {author} {\bibfnamefont {S.}~\bibnamefont {Eidelman}},
  \bibinfo {author} {\bibfnamefont {C.}~\bibnamefont {Hanhart}}, \bibinfo
  {author} {\bibfnamefont {A.}~\bibnamefont {Nefediev}}, \bibinfo {author}
  {\bibfnamefont {C.-P.}\ \bibnamefont {Shen}}, \bibinfo {author}
  {\bibfnamefont {C.~E.}\ \bibnamefont {Thomas}}, \bibinfo {author}
  {\bibfnamefont {A.}~\bibnamefont {Vairo}}, \ and\ \bibinfo {author}
  {\bibfnamefont {C.-Z.}\ \bibnamefont {Yuan}},\ }\href {\doibase
  10.1016/j.physrep.2020.05.001} {\bibfield  {journal} {\bibinfo  {journal}
  {Phys. Rept.}\ }\textbf {\bibinfo {volume} {873}},\ \bibinfo {pages} {1}
  (\bibinfo {year} {2020})},\ \Eprint {http://arxiv.org/abs/1907.07583}
  {arXiv:1907.07583 [hep-ex]} \BibitemShut {NoStop}%
\bibitem [{\citenamefont {Guo}\ \emph {et~al.}(2020)\citenamefont {Guo},
  \citenamefont {Liu},\ and\ \citenamefont {Sakai}}]{Guo:2019twa}%
  \BibitemOpen
  \bibfield  {author} {\bibinfo {author} {\bibfnamefont {F.-K.}\ \bibnamefont
  {Guo}}, \bibinfo {author} {\bibfnamefont {X.-H.}\ \bibnamefont {Liu}}, \ and\
  \bibinfo {author} {\bibfnamefont {S.}~\bibnamefont {Sakai}},\ }\href
  {\doibase 10.1016/j.ppnp.2020.103757} {\bibfield  {journal} {\bibinfo
  {journal} {Prog. Part. Nucl. Phys.}\ }\textbf {\bibinfo {volume} {112}},\
  \bibinfo {pages} {103757} (\bibinfo {year} {2020})},\ \Eprint
  {http://arxiv.org/abs/1912.07030} {arXiv:1912.07030 [hep-ph]} \BibitemShut
  {NoStop}%
\bibitem [{\citenamefont {Yang}\ \emph {et~al.}(2020)\citenamefont {Yang},
  \citenamefont {Ping},\ and\ \citenamefont {Segovia}}]{Yang:2020atz}%
  \BibitemOpen
  \bibfield  {author} {\bibinfo {author} {\bibfnamefont {G.}~\bibnamefont
  {Yang}}, \bibinfo {author} {\bibfnamefont {J.}~\bibnamefont {Ping}}, \ and\
  \bibinfo {author} {\bibfnamefont {J.}~\bibnamefont {Segovia}},\ }\href
  {\doibase 10.3390/sym12111869} {\bibfield  {journal} {\bibinfo  {journal}
  {Symmetry}\ }\textbf {\bibinfo {volume} {12}},\ \bibinfo {pages} {1869}
  (\bibinfo {year} {2020})},\ \Eprint {http://arxiv.org/abs/2009.00238}
  {arXiv:2009.00238 [hep-ph]} \BibitemShut {NoStop}%
\bibitem [{\citenamefont {Meng}\ \emph {et~al.}(2023)\citenamefont {Meng},
  \citenamefont {Wang}, \citenamefont {Wang},\ and\ \citenamefont
  {Zhu}}]{Meng:2022ozq}%
  \BibitemOpen
  \bibfield  {author} {\bibinfo {author} {\bibfnamefont {L.}~\bibnamefont
  {Meng}}, \bibinfo {author} {\bibfnamefont {B.}~\bibnamefont {Wang}}, \bibinfo
  {author} {\bibfnamefont {G.-J.}\ \bibnamefont {Wang}}, \ and\ \bibinfo
  {author} {\bibfnamefont {S.-L.}\ \bibnamefont {Zhu}},\ }\href {\doibase
  10.1016/j.physrep.2023.04.003} {\bibfield  {journal} {\bibinfo  {journal}
  {Phys. Rept.}\ }\textbf {\bibinfo {volume} {1019}},\ \bibinfo {pages} {1}
  (\bibinfo {year} {2023})},\ \Eprint {http://arxiv.org/abs/2204.08716}
  {arXiv:2204.08716 [hep-ph]} \BibitemShut {NoStop}%
\bibitem [{\citenamefont {Liu}\ \emph {et~al.}(2025)\citenamefont {Liu},
  \citenamefont {Pan}, \citenamefont {Liu}, \citenamefont {Wu}, \citenamefont
  {Lu},\ and\ \citenamefont {Geng}}]{Liu:2024uxn}%
  \BibitemOpen
  \bibfield  {author} {\bibinfo {author} {\bibfnamefont {M.-Z.}\ \bibnamefont
  {Liu}}, \bibinfo {author} {\bibfnamefont {Y.-W.}\ \bibnamefont {Pan}},
  \bibinfo {author} {\bibfnamefont {Z.-W.}\ \bibnamefont {Liu}}, \bibinfo
  {author} {\bibfnamefont {T.-W.}\ \bibnamefont {Wu}}, \bibinfo {author}
  {\bibfnamefont {J.-X.}\ \bibnamefont {Lu}}, \ and\ \bibinfo {author}
  {\bibfnamefont {L.-S.}\ \bibnamefont {Geng}},\ }\href {\doibase
  10.1016/j.physrep.2024.12.001} {\bibfield  {journal} {\bibinfo  {journal}
  {Phys. Rept.}\ }\textbf {\bibinfo {volume} {1108}},\ \bibinfo {pages} {1}
  (\bibinfo {year} {2025})},\ \Eprint {http://arxiv.org/abs/2404.06399}
  {arXiv:2404.06399 [hep-ph]} \BibitemShut {NoStop}%
\bibitem [{\citenamefont {Choi}\ \emph {et~al.}(2003)\citenamefont {Choi} \emph
  {et~al.}}]{Belle:2003nnu}%
  \BibitemOpen
  \bibfield  {author} {\bibinfo {author} {\bibfnamefont {S.~K.}\ \bibnamefont
  {Choi}} \emph {et~al.} (\bibinfo {collaboration} {Belle}),\ }\href {\doibase
  10.1103/PhysRevLett.91.262001} {\bibfield  {journal} {\bibinfo  {journal}
  {Phys. Rev. Lett.}\ }\textbf {\bibinfo {volume} {91}},\ \bibinfo {pages}
  {262001} (\bibinfo {year} {2003})},\ \Eprint
  {http://arxiv.org/abs/hep-ex/0309032} {arXiv:hep-ex/0309032} \BibitemShut
  {NoStop}%
\bibitem [{\citenamefont {Workman}\ \emph {et~al.}(2022)\citenamefont {Workman}
  \emph {et~al.}}]{ParticleDataGroup:2022pth}%
  \BibitemOpen
  \bibfield  {author} {\bibinfo {author} {\bibfnamefont {R.~L.}\ \bibnamefont
  {Workman}} \emph {et~al.} (\bibinfo {collaboration} {Particle Data Group}),\
  }\href {\doibase 10.1093/ptep/ptac097} {\bibfield  {journal} {\bibinfo
  {journal} {PTEP}\ }\textbf {\bibinfo {volume} {2022}},\ \bibinfo {pages}
  {083C01} (\bibinfo {year} {2022})}\BibitemShut {NoStop}%
\bibitem [{\citenamefont {Guo}\ and\ \citenamefont
  {Meissner}(2012)}]{Guo:2012tv}%
  \BibitemOpen
  \bibfield  {author} {\bibinfo {author} {\bibfnamefont {F.-K.}\ \bibnamefont
  {Guo}}\ and\ \bibinfo {author} {\bibfnamefont {U.-G.}\ \bibnamefont
  {Meissner}},\ }\href {\doibase 10.1103/PhysRevD.86.091501} {\bibfield
  {journal} {\bibinfo  {journal} {Phys. Rev. D}\ }\textbf {\bibinfo {volume}
  {86}},\ \bibinfo {pages} {091501} (\bibinfo {year} {2012})},\ \Eprint
  {http://arxiv.org/abs/1208.1134} {arXiv:1208.1134 [hep-ph]} \BibitemShut
  {NoStop}%
\bibitem [{\citenamefont {Olsen}(2015)}]{Olsen:2014maa}%
  \BibitemOpen
  \bibfield  {author} {\bibinfo {author} {\bibfnamefont {S.~L.}\ \bibnamefont
  {Olsen}},\ }\href {\doibase 10.1103/PhysRevD.91.057501} {\bibfield  {journal}
  {\bibinfo  {journal} {Phys. Rev. D}\ }\textbf {\bibinfo {volume} {91}},\
  \bibinfo {pages} {057501} (\bibinfo {year} {2015})},\ \Eprint
  {http://arxiv.org/abs/1410.6534} {arXiv:1410.6534 [hep-ex]} \BibitemShut
  {NoStop}%
\bibitem [{\citenamefont {Ablikim}\ \emph {et~al.}(2024)\citenamefont {Ablikim}
  \emph {et~al.}}]{BESIII:2023hml}%
  \BibitemOpen
  \bibfield  {author} {\bibinfo {author} {\bibfnamefont {M.}~\bibnamefont
  {Ablikim}} \emph {et~al.} (\bibinfo {collaboration} {BESIII}),\ }\href
  {\doibase 10.1103/PhysRevLett.132.151903} {\bibfield  {journal} {\bibinfo
  {journal} {Phys. Rev. Lett.}\ }\textbf {\bibinfo {volume} {132}},\ \bibinfo
  {pages} {151903} (\bibinfo {year} {2024})},\ \Eprint
  {http://arxiv.org/abs/2309.01502} {arXiv:2309.01502 [hep-ex]} \BibitemShut
  {NoStop}%
\bibitem [{\citenamefont {Aaij}\ \emph {et~al.}(2020)\citenamefont {Aaij} \emph
  {et~al.}}]{LHCb:2020xds}%
  \BibitemOpen
  \bibfield  {author} {\bibinfo {author} {\bibfnamefont {R.}~\bibnamefont
  {Aaij}} \emph {et~al.} (\bibinfo {collaboration} {LHCb}),\ }\href {\doibase
  10.1103/PhysRevD.102.092005} {\bibfield  {journal} {\bibinfo  {journal}
  {Phys. Rev. D}\ }\textbf {\bibinfo {volume} {102}},\ \bibinfo {pages}
  {092005} (\bibinfo {year} {2020})},\ \Eprint
  {http://arxiv.org/abs/2005.13419} {arXiv:2005.13419 [hep-ex]} \BibitemShut
  {NoStop}%
\bibitem [{\citenamefont {Aubert}\ \emph {et~al.}(2005)\citenamefont {Aubert}
  \emph {et~al.}}]{BaBar:2005hhc}%
  \BibitemOpen
  \bibfield  {author} {\bibinfo {author} {\bibfnamefont {B.}~\bibnamefont
  {Aubert}} \emph {et~al.} (\bibinfo {collaboration} {BaBar}),\ }\href
  {\doibase 10.1103/PhysRevLett.95.142001} {\bibfield  {journal} {\bibinfo
  {journal} {Phys. Rev. Lett.}\ }\textbf {\bibinfo {volume} {95}},\ \bibinfo
  {pages} {142001} (\bibinfo {year} {2005})},\ \Eprint
  {http://arxiv.org/abs/hep-ex/0506081} {arXiv:hep-ex/0506081} \BibitemShut
  {NoStop}%
\bibitem [{\citenamefont {Coan}\ \emph {et~al.}(2006)\citenamefont {Coan} \emph
  {et~al.}}]{CLEO:2006ike}%
  \BibitemOpen
  \bibfield  {author} {\bibinfo {author} {\bibfnamefont {T.~E.}\ \bibnamefont
  {Coan}} \emph {et~al.} (\bibinfo {collaboration} {CLEO}),\ }\href {\doibase
  10.1103/PhysRevLett.96.162003} {\bibfield  {journal} {\bibinfo  {journal}
  {Phys. Rev. Lett.}\ }\textbf {\bibinfo {volume} {96}},\ \bibinfo {pages}
  {162003} (\bibinfo {year} {2006})},\ \Eprint
  {http://arxiv.org/abs/hep-ex/0602034} {arXiv:hep-ex/0602034} \BibitemShut
  {NoStop}%
\bibitem [{\citenamefont {Yuan}\ \emph {et~al.}(2007)\citenamefont {Yuan} \emph
  {et~al.}}]{Belle:2007dxy}%
  \BibitemOpen
  \bibfield  {author} {\bibinfo {author} {\bibfnamefont {C.~Z.}\ \bibnamefont
  {Yuan}} \emph {et~al.} (\bibinfo {collaboration} {Belle}),\ }\href {\doibase
  10.1103/PhysRevLett.99.182004} {\bibfield  {journal} {\bibinfo  {journal}
  {Phys. Rev. Lett.}\ }\textbf {\bibinfo {volume} {99}},\ \bibinfo {pages}
  {182004} (\bibinfo {year} {2007})},\ \Eprint {http://arxiv.org/abs/0707.2541}
  {arXiv:0707.2541 [hep-ex]} \BibitemShut {NoStop}%
\bibitem [{\citenamefont {Ablikim}\ \emph {et~al.}(2007)\citenamefont {Ablikim}
  \emph {et~al.}}]{BES:2007zwq}%
  \BibitemOpen
  \bibfield  {author} {\bibinfo {author} {\bibfnamefont {M.}~\bibnamefont
  {Ablikim}} \emph {et~al.} (\bibinfo {collaboration} {BES}),\ }\href {\doibase
  10.1016/j.physletb.2007.11.100} {\bibfield  {journal} {\bibinfo  {journal}
  {eConf}\ }\textbf {\bibinfo {volume} {C070805}},\ \bibinfo {pages} {02}
  (\bibinfo {year} {2007})},\ \Eprint {http://arxiv.org/abs/0705.4500}
  {arXiv:0705.4500 [hep-ex]} \BibitemShut {NoStop}%
\bibitem [{\citenamefont {Ablikim}\ \emph
  {et~al.}(2017{\natexlab{a}})\citenamefont {Ablikim} \emph
  {et~al.}}]{BESIII:2016bnd}%
  \BibitemOpen
  \bibfield  {author} {\bibinfo {author} {\bibfnamefont {M.}~\bibnamefont
  {Ablikim}} \emph {et~al.} (\bibinfo {collaboration} {BESIII}),\ }\href
  {\doibase 10.1103/PhysRevLett.118.092001} {\bibfield  {journal} {\bibinfo
  {journal} {Phys. Rev. Lett.}\ }\textbf {\bibinfo {volume} {118}},\ \bibinfo
  {pages} {092001} (\bibinfo {year} {2017}{\natexlab{a}})},\ \Eprint
  {http://arxiv.org/abs/1611.01317} {arXiv:1611.01317 [hep-ex]} \BibitemShut
  {NoStop}%
\bibitem [{\citenamefont {Ablikim}\ \emph
  {et~al.}(2017{\natexlab{b}})\citenamefont {Ablikim} \emph
  {et~al.}}]{BESIII:2016adj}%
  \BibitemOpen
  \bibfield  {author} {\bibinfo {author} {\bibfnamefont {M.}~\bibnamefont
  {Ablikim}} \emph {et~al.} (\bibinfo {collaboration} {BESIII}),\ }\href
  {\doibase 10.1103/PhysRevLett.118.092002} {\bibfield  {journal} {\bibinfo
  {journal} {Phys. Rev. Lett.}\ }\textbf {\bibinfo {volume} {118}},\ \bibinfo
  {pages} {092002} (\bibinfo {year} {2017}{\natexlab{b}})},\ \Eprint
  {http://arxiv.org/abs/1610.07044} {arXiv:1610.07044 [hep-ex]} \BibitemShut
  {NoStop}%
\bibitem [{\citenamefont {Ablikim}\ \emph
  {et~al.}(2017{\natexlab{c}})\citenamefont {Ablikim} \emph
  {et~al.}}]{BESIII:2017tqk}%
  \BibitemOpen
  \bibfield  {author} {\bibinfo {author} {\bibfnamefont {M.}~\bibnamefont
  {Ablikim}} \emph {et~al.} (\bibinfo {collaboration} {BESIII}),\ }\href
  {\doibase 10.1103/PhysRevD.96.032004} {\bibfield  {journal} {\bibinfo
  {journal} {Phys. Rev. D}\ }\textbf {\bibinfo {volume} {96}},\ \bibinfo
  {pages} {032004} (\bibinfo {year} {2017}{\natexlab{c}})},\ \bibinfo {note}
  {[Erratum: Phys.Rev.D 99, 019903 (2019)]},\ \Eprint
  {http://arxiv.org/abs/1703.08787} {arXiv:1703.08787 [hep-ex]} \BibitemShut
  {NoStop}%
\bibitem [{\citenamefont {Ablikim}\ \emph {et~al.}(2021)\citenamefont {Ablikim}
  \emph {et~al.}}]{BESIII:2021njb}%
  \BibitemOpen
  \bibfield  {author} {\bibinfo {author} {\bibfnamefont {M.}~\bibnamefont
  {Ablikim}} \emph {et~al.} (\bibinfo {collaboration} {BESIII}),\ }\href
  {\doibase 10.1103/PhysRevD.104.052012} {\bibfield  {journal} {\bibinfo
  {journal} {Phys. Rev. D}\ }\textbf {\bibinfo {volume} {104}},\ \bibinfo
  {pages} {052012} (\bibinfo {year} {2021})},\ \Eprint
  {http://arxiv.org/abs/2107.09210} {arXiv:2107.09210 [hep-ex]} \BibitemShut
  {NoStop}%
\bibitem [{\citenamefont {Ablikim}\ \emph {et~al.}(2020)\citenamefont {Ablikim}
  \emph {et~al.}}]{BESIII:2020bgb}%
  \BibitemOpen
  \bibfield  {author} {\bibinfo {author} {\bibfnamefont {M.}~\bibnamefont
  {Ablikim}} \emph {et~al.} (\bibinfo {collaboration} {BESIII}),\ }\href
  {\doibase 10.1103/PhysRevD.102.031101} {\bibfield  {journal} {\bibinfo
  {journal} {Phys. Rev. D}\ }\textbf {\bibinfo {volume} {102}},\ \bibinfo
  {pages} {031101} (\bibinfo {year} {2020})},\ \Eprint
  {http://arxiv.org/abs/2003.03705} {arXiv:2003.03705 [hep-ex]} \BibitemShut
  {NoStop}%
\bibitem [{\citenamefont {Chen}\ \emph
  {et~al.}(2016{\natexlab{b}})\citenamefont {Chen}, \citenamefont {Liu},
  \citenamefont {Li},\ and\ \citenamefont {Ke}}]{Chen:2015bft}%
  \BibitemOpen
  \bibfield  {author} {\bibinfo {author} {\bibfnamefont {D.-Y.}\ \bibnamefont
  {Chen}}, \bibinfo {author} {\bibfnamefont {X.}~\bibnamefont {Liu}}, \bibinfo
  {author} {\bibfnamefont {X.-Q.}\ \bibnamefont {Li}}, \ and\ \bibinfo {author}
  {\bibfnamefont {H.-W.}\ \bibnamefont {Ke}},\ }\href {\doibase
  10.1103/PhysRevD.93.014011} {\bibfield  {journal} {\bibinfo  {journal} {Phys.
  Rev. D}\ }\textbf {\bibinfo {volume} {93}},\ \bibinfo {pages} {014011}
  (\bibinfo {year} {2016}{\natexlab{b}})},\ \Eprint
  {http://arxiv.org/abs/1512.04157} {arXiv:1512.04157 [hep-ph]} \BibitemShut
  {NoStop}%
\bibitem [{\citenamefont {Fu}\ and\ \citenamefont {Jiang}(2019)}]{Fu:2018yxq}%
  \BibitemOpen
  \bibfield  {author} {\bibinfo {author} {\bibfnamefont {H.-F.}\ \bibnamefont
  {Fu}}\ and\ \bibinfo {author} {\bibfnamefont {L.}~\bibnamefont {Jiang}},\
  }\href {\doibase 10.1140/epjc/s10052-019-6976-0} {\bibfield  {journal}
  {\bibinfo  {journal} {Eur. Phys. J. C}\ }\textbf {\bibinfo {volume} {79}},\
  \bibinfo {pages} {460} (\bibinfo {year} {2019})},\ \Eprint
  {http://arxiv.org/abs/1812.00179} {arXiv:1812.00179 [hep-ph]} \BibitemShut
  {NoStop}%
\bibitem [{\citenamefont {Wang}\ \emph {et~al.}(2019)\citenamefont {Wang},
  \citenamefont {Chen}, \citenamefont {Liu},\ and\ \citenamefont
  {Matsuki}}]{Wang:2019mhs}%
  \BibitemOpen
  \bibfield  {author} {\bibinfo {author} {\bibfnamefont {J.-Z.}\ \bibnamefont
  {Wang}}, \bibinfo {author} {\bibfnamefont {D.-Y.}\ \bibnamefont {Chen}},
  \bibinfo {author} {\bibfnamefont {X.}~\bibnamefont {Liu}}, \ and\ \bibinfo
  {author} {\bibfnamefont {T.}~\bibnamefont {Matsuki}},\ }\href {\doibase
  10.1103/PhysRevD.99.114003} {\bibfield  {journal} {\bibinfo  {journal} {Phys.
  Rev. D}\ }\textbf {\bibinfo {volume} {99}},\ \bibinfo {pages} {114003}
  (\bibinfo {year} {2019})},\ \Eprint {http://arxiv.org/abs/1903.07115}
  {arXiv:1903.07115 [hep-ph]} \BibitemShut {NoStop}%
\bibitem [{\citenamefont {Wang}(2021)}]{Wang:2021qus}%
  \BibitemOpen
  \bibfield  {author} {\bibinfo {author} {\bibfnamefont {Z.-G.}\ \bibnamefont
  {Wang}},\ }\href {\doibase 10.1016/j.nuclphysb.2021.115592} {\bibfield
  {journal} {\bibinfo  {journal} {Nucl. Phys. B}\ }\textbf {\bibinfo {volume}
  {973}},\ \bibinfo {pages} {115592} (\bibinfo {year} {2021})},\ \Eprint
  {http://arxiv.org/abs/2108.05759} {arXiv:2108.05759 [hep-ph]} \BibitemShut
  {NoStop}%
\bibitem [{\citenamefont {Zhou}\ \emph {et~al.}(2023)\citenamefont {Zhou},
  \citenamefont {Li},\ and\ \citenamefont {Xiao}}]{Zhou:2023yjv}%
  \BibitemOpen
  \bibfield  {author} {\bibinfo {author} {\bibfnamefont {Z.-Y.}\ \bibnamefont
  {Zhou}}, \bibinfo {author} {\bibfnamefont {C.-Y.}\ \bibnamefont {Li}}, \ and\
  \bibinfo {author} {\bibfnamefont {Z.}~\bibnamefont {Xiao}},\ }\href@noop {}
  {\  (\bibinfo {year} {2023})},\ \Eprint {http://arxiv.org/abs/2304.07052}
  {arXiv:2304.07052 [hep-ph]} \BibitemShut {NoStop}%
\bibitem [{\citenamefont {Nakamura}\ \emph {et~al.}(2023)\citenamefont
  {Nakamura}, \citenamefont {Li}, \citenamefont {Peng}, \citenamefont {Sun},\
  and\ \citenamefont {Zhou}}]{Nakamura:2023obk}%
  \BibitemOpen
  \bibfield  {author} {\bibinfo {author} {\bibfnamefont {S.~X.}\ \bibnamefont
  {Nakamura}}, \bibinfo {author} {\bibfnamefont {X.~H.}\ \bibnamefont {Li}},
  \bibinfo {author} {\bibfnamefont {H.~P.}\ \bibnamefont {Peng}}, \bibinfo
  {author} {\bibfnamefont {Z.~T.}\ \bibnamefont {Sun}}, \ and\ \bibinfo
  {author} {\bibfnamefont {X.~R.}\ \bibnamefont {Zhou}},\ }\href@noop {} {\
  (\bibinfo {year} {2023})},\ \Eprint {http://arxiv.org/abs/2312.17658}
  {arXiv:2312.17658 [hep-ph]} \BibitemShut {NoStop}%
\bibitem [{\citenamefont {von Detten}\ \emph {et~al.}(2024)\citenamefont {von
  Detten}, \citenamefont {Baru}, \citenamefont {Hanhart}, \citenamefont {Wang},
  \citenamefont {Winney},\ and\ \citenamefont {Zhao}}]{vonDetten:2024eie}%
  \BibitemOpen
  \bibfield  {author} {\bibinfo {author} {\bibfnamefont {L.}~\bibnamefont {von
  Detten}}, \bibinfo {author} {\bibfnamefont {V.}~\bibnamefont {Baru}},
  \bibinfo {author} {\bibfnamefont {C.}~\bibnamefont {Hanhart}}, \bibinfo
  {author} {\bibfnamefont {Q.}~\bibnamefont {Wang}}, \bibinfo {author}
  {\bibfnamefont {D.}~\bibnamefont {Winney}}, \ and\ \bibinfo {author}
  {\bibfnamefont {Q.}~\bibnamefont {Zhao}},\ }\href {\doibase
  10.1103/PhysRevD.109.116002} {\bibfield  {journal} {\bibinfo  {journal}
  {Phys. Rev. D}\ }\textbf {\bibinfo {volume} {109}},\ \bibinfo {pages}
  {116002} (\bibinfo {year} {2024})},\ \Eprint
  {http://arxiv.org/abs/2402.03057} {arXiv:2402.03057 [hep-ph]} \BibitemShut
  {NoStop}%
\bibitem [{\citenamefont {Wang}(2017)}]{Wang:2016wwe}%
  \BibitemOpen
  \bibfield  {author} {\bibinfo {author} {\bibfnamefont {Z.-G.}\ \bibnamefont
  {Wang}},\ }\href {\doibase 10.1088/1674-1137/41/8/083103} {\bibfield
  {journal} {\bibinfo  {journal} {Chin. Phys. C}\ }\textbf {\bibinfo {volume}
  {41}},\ \bibinfo {pages} {083103} (\bibinfo {year} {2017})},\ \Eprint
  {http://arxiv.org/abs/1611.03250} {arXiv:1611.03250 [hep-ph]} \BibitemShut
  {NoStop}%
\bibitem [{\citenamefont {Anwar}\ and\ \citenamefont
  {Lu}(2021)}]{Anwar:2021dmg}%
  \BibitemOpen
  \bibfield  {author} {\bibinfo {author} {\bibfnamefont {M.~N.}\ \bibnamefont
  {Anwar}}\ and\ \bibinfo {author} {\bibfnamefont {Y.}~\bibnamefont {Lu}},\
  }\href {\doibase 10.1103/PhysRevD.104.094006} {\bibfield  {journal} {\bibinfo
   {journal} {Phys. Rev. D}\ }\textbf {\bibinfo {volume} {104}},\ \bibinfo
  {pages} {094006} (\bibinfo {year} {2021})},\ \Eprint
  {http://arxiv.org/abs/2109.02539} {arXiv:2109.02539 [hep-ph]} \BibitemShut
  {NoStop}%
\bibitem [{\citenamefont {Peng}\ \emph {et~al.}(2023)\citenamefont {Peng},
  \citenamefont {Yan}, \citenamefont {S\'anchez~S\'anchez},\ and\ \citenamefont
  {Pavon~Valderrama}}]{Peng:2022nrj}%
  \BibitemOpen
  \bibfield  {author} {\bibinfo {author} {\bibfnamefont {F.-Z.}\ \bibnamefont
  {Peng}}, \bibinfo {author} {\bibfnamefont {M.-J.}\ \bibnamefont {Yan}},
  \bibinfo {author} {\bibfnamefont {M.}~\bibnamefont {S\'anchez~S\'anchez}}, \
  and\ \bibinfo {author} {\bibfnamefont {M.}~\bibnamefont {Pavon~Valderrama}},\
  }\href {\doibase 10.1103/PhysRevD.107.016001} {\bibfield  {journal} {\bibinfo
   {journal} {Phys. Rev. D}\ }\textbf {\bibinfo {volume} {107}},\ \bibinfo
  {pages} {016001} (\bibinfo {year} {2023})},\ \Eprint
  {http://arxiv.org/abs/2205.13590} {arXiv:2205.13590 [hep-ph]} \BibitemShut
  {NoStop}%
\bibitem [{\citenamefont {Ji}\ \emph {et~al.}(2022)\citenamefont {Ji},
  \citenamefont {Dong}, \citenamefont {Guo},\ and\ \citenamefont
  {Zou}}]{Ji:2022blw}%
  \BibitemOpen
  \bibfield  {author} {\bibinfo {author} {\bibfnamefont {T.}~\bibnamefont
  {Ji}}, \bibinfo {author} {\bibfnamefont {X.-K.}\ \bibnamefont {Dong}},
  \bibinfo {author} {\bibfnamefont {F.-K.}\ \bibnamefont {Guo}}, \ and\
  \bibinfo {author} {\bibfnamefont {B.-S.}\ \bibnamefont {Zou}},\ }\href
  {\doibase 10.1103/PhysRevLett.129.102002} {\bibfield  {journal} {\bibinfo
  {journal} {Phys. Rev. Lett.}\ }\textbf {\bibinfo {volume} {129}},\ \bibinfo
  {pages} {102002} (\bibinfo {year} {2022})},\ \Eprint
  {http://arxiv.org/abs/2205.10994} {arXiv:2205.10994 [hep-ph]} \BibitemShut
  {NoStop}%
\bibitem [{\citenamefont {Wang}\ \emph {et~al.}(2024)\citenamefont {Wang},
  \citenamefont {Wang}, \citenamefont {Wang},\ and\ \citenamefont
  {Liu}}]{Wang:2023ivd}%
  \BibitemOpen
  \bibfield  {author} {\bibinfo {author} {\bibfnamefont {Z.-P.}\ \bibnamefont
  {Wang}}, \bibinfo {author} {\bibfnamefont {F.-L.}\ \bibnamefont {Wang}},
  \bibinfo {author} {\bibfnamefont {G.-J.}\ \bibnamefont {Wang}}, \ and\
  \bibinfo {author} {\bibfnamefont {X.}~\bibnamefont {Liu}},\ }\href {\doibase
  10.1103/PhysRevD.110.L051501} {\bibfield  {journal} {\bibinfo  {journal}
  {Phys. Rev. D}\ }\textbf {\bibinfo {volume} {110}},\ \bibinfo {pages}
  {L051501} (\bibinfo {year} {2024})},\ \Eprint
  {http://arxiv.org/abs/2312.03512} {arXiv:2312.03512 [hep-ph]} \BibitemShut
  {NoStop}%
\bibitem [{\citenamefont {Liu}\ \emph {et~al.}(2024{\natexlab{a}})\citenamefont
  {Liu}, \citenamefont {Ling},\ and\ \citenamefont {Geng}}]{Liu:2024ziu}%
  \BibitemOpen
  \bibfield  {author} {\bibinfo {author} {\bibfnamefont {M.-Z.}\ \bibnamefont
  {Liu}}, \bibinfo {author} {\bibfnamefont {X.-Z.}\ \bibnamefont {Ling}}, \
  and\ \bibinfo {author} {\bibfnamefont {L.-S.}\ \bibnamefont {Geng}},\ }\href
  {\doibase 10.1103/PhysRevD.110.054035} {\bibfield  {journal} {\bibinfo
  {journal} {Phys. Rev. D}\ }\textbf {\bibinfo {volume} {110}},\ \bibinfo
  {pages} {054035} (\bibinfo {year} {2024}{\natexlab{a}})},\ \Eprint
  {http://arxiv.org/abs/2404.07681} {arXiv:2404.07681 [hep-ph]} \BibitemShut
  {NoStop}%
\bibitem [{\citenamefont {Wang}\ \emph {et~al.}(2013)\citenamefont {Wang},
  \citenamefont {Hanhart},\ and\ \citenamefont {Zhao}}]{Wang:2013cya}%
  \BibitemOpen
  \bibfield  {author} {\bibinfo {author} {\bibfnamefont {Q.}~\bibnamefont
  {Wang}}, \bibinfo {author} {\bibfnamefont {C.}~\bibnamefont {Hanhart}}, \
  and\ \bibinfo {author} {\bibfnamefont {Q.}~\bibnamefont {Zhao}},\ }\href
  {\doibase 10.1103/PhysRevLett.111.132003} {\bibfield  {journal} {\bibinfo
  {journal} {Phys. Rev. Lett.}\ }\textbf {\bibinfo {volume} {111}},\ \bibinfo
  {pages} {132003} (\bibinfo {year} {2013})},\ \Eprint
  {http://arxiv.org/abs/1303.6355} {arXiv:1303.6355 [hep-ph]} \BibitemShut
  {NoStop}%
\bibitem [{\citenamefont {Dong}\ \emph
  {et~al.}(2014{\natexlab{a}})\citenamefont {Dong}, \citenamefont {Faessler},
  \citenamefont {Gutsche},\ and\ \citenamefont {Lyubovitskij}}]{Dong:2013kta}%
  \BibitemOpen
  \bibfield  {author} {\bibinfo {author} {\bibfnamefont {Y.}~\bibnamefont
  {Dong}}, \bibinfo {author} {\bibfnamefont {A.}~\bibnamefont {Faessler}},
  \bibinfo {author} {\bibfnamefont {T.}~\bibnamefont {Gutsche}}, \ and\
  \bibinfo {author} {\bibfnamefont {V.~E.}\ \bibnamefont {Lyubovitskij}},\
  }\href {\doibase 10.1103/PhysRevD.89.034018} {\bibfield  {journal} {\bibinfo
  {journal} {Phys. Rev. D}\ }\textbf {\bibinfo {volume} {89}},\ \bibinfo
  {pages} {034018} (\bibinfo {year} {2014}{\natexlab{a}})},\ \Eprint
  {http://arxiv.org/abs/1310.4373} {arXiv:1310.4373 [hep-ph]} \BibitemShut
  {NoStop}%
\bibitem [{\citenamefont {Liu}\ and\ \citenamefont {Li}(2013)}]{Liu:2013vfa}%
  \BibitemOpen
  \bibfield  {author} {\bibinfo {author} {\bibfnamefont {X.-H.}\ \bibnamefont
  {Liu}}\ and\ \bibinfo {author} {\bibfnamefont {G.}~\bibnamefont {Li}},\
  }\href {\doibase 10.1103/PhysRevD.88.014013} {\bibfield  {journal} {\bibinfo
  {journal} {Phys. Rev. D}\ }\textbf {\bibinfo {volume} {88}},\ \bibinfo
  {pages} {014013} (\bibinfo {year} {2013})},\ \Eprint
  {http://arxiv.org/abs/1306.1384} {arXiv:1306.1384 [hep-ph]} \BibitemShut
  {NoStop}%
\bibitem [{\citenamefont {Guo}\ \emph {et~al.}(2013)\citenamefont {Guo},
  \citenamefont {Hanhart}, \citenamefont {Mei\ss{}ner}, \citenamefont {Wang},\
  and\ \citenamefont {Zhao}}]{Guo:2013zbw}%
  \BibitemOpen
  \bibfield  {author} {\bibinfo {author} {\bibfnamefont {F.-K.}\ \bibnamefont
  {Guo}}, \bibinfo {author} {\bibfnamefont {C.}~\bibnamefont {Hanhart}},
  \bibinfo {author} {\bibfnamefont {U.-G.}\ \bibnamefont {Mei\ss{}ner}},
  \bibinfo {author} {\bibfnamefont {Q.}~\bibnamefont {Wang}}, \ and\ \bibinfo
  {author} {\bibfnamefont {Q.}~\bibnamefont {Zhao}},\ }\href {\doibase
  10.1016/j.physletb.2013.06.053} {\bibfield  {journal} {\bibinfo  {journal}
  {Phys. Lett. B}\ }\textbf {\bibinfo {volume} {725}},\ \bibinfo {pages} {127}
  (\bibinfo {year} {2013})},\ \Eprint {http://arxiv.org/abs/1306.3096}
  {arXiv:1306.3096 [hep-ph]} \BibitemShut {NoStop}%
\bibitem [{\citenamefont {Dong}\ \emph
  {et~al.}(2014{\natexlab{b}})\citenamefont {Dong}, \citenamefont {Faessler},
  \citenamefont {Gutsche},\ and\ \citenamefont {Lyubovitskij}}]{Dong:2014zka}%
  \BibitemOpen
  \bibfield  {author} {\bibinfo {author} {\bibfnamefont {Y.}~\bibnamefont
  {Dong}}, \bibinfo {author} {\bibfnamefont {A.}~\bibnamefont {Faessler}},
  \bibinfo {author} {\bibfnamefont {T.}~\bibnamefont {Gutsche}}, \ and\
  \bibinfo {author} {\bibfnamefont {V.~E.}\ \bibnamefont {Lyubovitskij}},\
  }\href {\doibase 10.1103/PhysRevD.90.074032} {\bibfield  {journal} {\bibinfo
  {journal} {Phys. Rev. D}\ }\textbf {\bibinfo {volume} {90}},\ \bibinfo
  {pages} {074032} (\bibinfo {year} {2014}{\natexlab{b}})},\ \Eprint
  {http://arxiv.org/abs/1404.6161} {arXiv:1404.6161 [hep-ph]} \BibitemShut
  {NoStop}%
\bibitem [{\citenamefont {Qin}\ \emph {et~al.}(2016)\citenamefont {Qin},
  \citenamefont {Xue},\ and\ \citenamefont {Zhao}}]{Qin:2016spb}%
  \BibitemOpen
  \bibfield  {author} {\bibinfo {author} {\bibfnamefont {W.}~\bibnamefont
  {Qin}}, \bibinfo {author} {\bibfnamefont {S.-R.}\ \bibnamefont {Xue}}, \ and\
  \bibinfo {author} {\bibfnamefont {Q.}~\bibnamefont {Zhao}},\ }\href {\doibase
  10.1103/PhysRevD.94.054035} {\bibfield  {journal} {\bibinfo  {journal} {Phys.
  Rev. D}\ }\textbf {\bibinfo {volume} {94}},\ \bibinfo {pages} {054035}
  (\bibinfo {year} {2016})},\ \Eprint {http://arxiv.org/abs/1605.02407}
  {arXiv:1605.02407 [hep-ph]} \BibitemShut {NoStop}%
\bibitem [{\citenamefont {Chen}\ \emph {et~al.}(2017)\citenamefont {Chen},
  \citenamefont {Xiao},\ and\ \citenamefont {He}}]{Chen:2017abq}%
  \BibitemOpen
  \bibfield  {author} {\bibinfo {author} {\bibfnamefont {D.-Y.}\ \bibnamefont
  {Chen}}, \bibinfo {author} {\bibfnamefont {C.-J.}\ \bibnamefont {Xiao}}, \
  and\ \bibinfo {author} {\bibfnamefont {J.}~\bibnamefont {He}},\ }\href
  {\doibase 10.1103/PhysRevD.96.054017} {\bibfield  {journal} {\bibinfo
  {journal} {Phys. Rev. D}\ }\textbf {\bibinfo {volume} {96}},\ \bibinfo
  {pages} {054017} (\bibinfo {year} {2017})}\BibitemShut {NoStop}%
\bibitem [{\citenamefont {Wang}(2023)}]{Wang:2023dsm}%
  \BibitemOpen
  \bibfield  {author} {\bibinfo {author} {\bibfnamefont {Z.-G.}\ \bibnamefont
  {Wang}},\ }\href {\doibase 10.1142/S0217751X23501750} {\bibfield  {journal}
  {\bibinfo  {journal} {Int. J. Mod. Phys. A}\ }\textbf {\bibinfo {volume}
  {38}},\ \bibinfo {pages} {2350175} (\bibinfo {year} {2023})},\ \Eprint
  {http://arxiv.org/abs/2309.01337} {arXiv:2309.01337 [hep-ph]} \BibitemShut
  {NoStop}%
\bibitem [{\citenamefont {Aubert}\ \emph {et~al.}(2006)\citenamefont {Aubert}
  \emph {et~al.}}]{BaBar:2005xmz}%
  \BibitemOpen
  \bibfield  {author} {\bibinfo {author} {\bibfnamefont {B.}~\bibnamefont
  {Aubert}} \emph {et~al.} (\bibinfo {collaboration} {BaBar}),\ }\href
  {\doibase 10.1103/PhysRevD.73.011101} {\bibfield  {journal} {\bibinfo
  {journal} {Phys. Rev. D}\ }\textbf {\bibinfo {volume} {73}},\ \bibinfo
  {pages} {011101} (\bibinfo {year} {2006})},\ \Eprint
  {http://arxiv.org/abs/hep-ex/0507090} {arXiv:hep-ex/0507090} \BibitemShut
  {NoStop}%
\bibitem [{\citenamefont {Garg}\ \emph {et~al.}(2019)\citenamefont {Garg} \emph
  {et~al.}}]{Belle:2019pfg}%
  \BibitemOpen
  \bibfield  {author} {\bibinfo {author} {\bibfnamefont {R.}~\bibnamefont
  {Garg}} \emph {et~al.} (\bibinfo {collaboration} {Belle}),\ }\href {\doibase
  10.1103/PhysRevD.99.071102} {\bibfield  {journal} {\bibinfo  {journal} {Phys.
  Rev. D}\ }\textbf {\bibinfo {volume} {99}},\ \bibinfo {pages} {071102}
  (\bibinfo {year} {2019})},\ \Eprint {http://arxiv.org/abs/1901.06470}
  {arXiv:1901.06470 [hep-ex]} \BibitemShut {NoStop}%
\bibitem [{\citenamefont {Aaij}\ \emph {et~al.}(2022)\citenamefont {Aaij} \emph
  {et~al.}}]{LHCb:2022oqs}%
  \BibitemOpen
  \bibfield  {author} {\bibinfo {author} {\bibfnamefont {R.}~\bibnamefont
  {Aaij}} \emph {et~al.} (\bibinfo {collaboration} {LHCb}),\ }\href {\doibase
  10.1007/JHEP04(2022)046} {\bibfield  {journal} {\bibinfo  {journal} {JHEP}\
  }\textbf {\bibinfo {volume} {04}},\ \bibinfo {pages} {046} (\bibinfo {year}
  {2022})},\ \Eprint {http://arxiv.org/abs/2202.04045} {arXiv:2202.04045
  [hep-ex]} \BibitemShut {NoStop}%
\bibitem [{\citenamefont {Albuquerque}\ \emph {et~al.}(2015)\citenamefont
  {Albuquerque}, \citenamefont {Nielsen},\ and\ \citenamefont
  {Zanetti}}]{Albuquerque:2015nwa}%
  \BibitemOpen
  \bibfield  {author} {\bibinfo {author} {\bibfnamefont {R.~M.}\ \bibnamefont
  {Albuquerque}}, \bibinfo {author} {\bibfnamefont {M.}~\bibnamefont
  {Nielsen}}, \ and\ \bibinfo {author} {\bibfnamefont {C.~M.}\ \bibnamefont
  {Zanetti}},\ }\href {\doibase 10.1016/j.physletb.2015.05.022} {\bibfield
  {journal} {\bibinfo  {journal} {Phys. Lett. B}\ }\textbf {\bibinfo {volume}
  {747}},\ \bibinfo {pages} {83} (\bibinfo {year} {2015})},\ \Eprint
  {http://arxiv.org/abs/1502.00119} {arXiv:1502.00119 [hep-ph]} \BibitemShut
  {NoStop}%
\bibitem [{\citenamefont {Chen}\ \emph {et~al.}(2021)\citenamefont {Chen},
  \citenamefont {Han}, \citenamefont {L\"u}, \citenamefont {Wang},\ and\
  \citenamefont {Yu}}]{Chen:2020eyu}%
  \BibitemOpen
  \bibfield  {author} {\bibinfo {author} {\bibfnamefont {Y.-K.}\ \bibnamefont
  {Chen}}, \bibinfo {author} {\bibfnamefont {J.-J.}\ \bibnamefont {Han}},
  \bibinfo {author} {\bibfnamefont {Q.-F.}\ \bibnamefont {L\"u}}, \bibinfo
  {author} {\bibfnamefont {J.-P.}\ \bibnamefont {Wang}}, \ and\ \bibinfo
  {author} {\bibfnamefont {F.-S.}\ \bibnamefont {Yu}},\ }\href {\doibase
  10.1140/epjc/s10052-021-08857-8} {\bibfield  {journal} {\bibinfo  {journal}
  {Eur. Phys. J. C}\ }\textbf {\bibinfo {volume} {81}},\ \bibinfo {pages} {71}
  (\bibinfo {year} {2021})},\ \Eprint {http://arxiv.org/abs/2009.01182}
  {arXiv:2009.01182 [hep-ph]} \BibitemShut {NoStop}%
\bibitem [{\citenamefont {Cheng}\ \emph {et~al.}(2005)\citenamefont {Cheng},
  \citenamefont {Chua},\ and\ \citenamefont {Soni}}]{Cheng:2004ru}%
  \BibitemOpen
  \bibfield  {author} {\bibinfo {author} {\bibfnamefont {H.-Y.}\ \bibnamefont
  {Cheng}}, \bibinfo {author} {\bibfnamefont {C.-K.}\ \bibnamefont {Chua}}, \
  and\ \bibinfo {author} {\bibfnamefont {A.}~\bibnamefont {Soni}},\ }\href
  {\doibase 10.1103/PhysRevD.71.014030} {\bibfield  {journal} {\bibinfo
  {journal} {Phys. Rev. D}\ }\textbf {\bibinfo {volume} {71}},\ \bibinfo
  {pages} {014030} (\bibinfo {year} {2005})},\ \Eprint
  {http://arxiv.org/abs/hep-ph/0409317} {arXiv:hep-ph/0409317} \BibitemShut
  {NoStop}%
\bibitem [{\citenamefont {Yu}\ \emph {et~al.}(2018)\citenamefont {Yu},
  \citenamefont {Jiang}, \citenamefont {Li}, \citenamefont {L\"u},
  \citenamefont {Wang},\ and\ \citenamefont {Zhao}}]{Yu:2017zst}%
  \BibitemOpen
  \bibfield  {author} {\bibinfo {author} {\bibfnamefont {F.-S.}\ \bibnamefont
  {Yu}}, \bibinfo {author} {\bibfnamefont {H.-Y.}\ \bibnamefont {Jiang}},
  \bibinfo {author} {\bibfnamefont {R.-H.}\ \bibnamefont {Li}}, \bibinfo
  {author} {\bibfnamefont {C.-D.}\ \bibnamefont {L\"u}}, \bibinfo {author}
  {\bibfnamefont {W.}~\bibnamefont {Wang}}, \ and\ \bibinfo {author}
  {\bibfnamefont {Z.-X.}\ \bibnamefont {Zhao}},\ }\href {\doibase
  10.1088/1674-1137/42/5/051001} {\bibfield  {journal} {\bibinfo  {journal}
  {Chin. Phys. C}\ }\textbf {\bibinfo {volume} {42}},\ \bibinfo {pages}
  {051001} (\bibinfo {year} {2018})},\ \Eprint
  {http://arxiv.org/abs/1703.09086} {arXiv:1703.09086 [hep-ph]} \BibitemShut
  {NoStop}%
\bibitem [{\citenamefont {Liu}\ \emph {et~al.}(2007)\citenamefont {Liu},
  \citenamefont {Zhang},\ and\ \citenamefont {Zhu}}]{Liu:2006df}%
  \BibitemOpen
  \bibfield  {author} {\bibinfo {author} {\bibfnamefont {X.}~\bibnamefont
  {Liu}}, \bibinfo {author} {\bibfnamefont {B.}~\bibnamefont {Zhang}}, \ and\
  \bibinfo {author} {\bibfnamefont {S.-L.}\ \bibnamefont {Zhu}},\ }\href
  {\doibase 10.1016/j.physletb.2006.12.031} {\bibfield  {journal} {\bibinfo
  {journal} {Phys. Lett. B}\ }\textbf {\bibinfo {volume} {645}},\ \bibinfo
  {pages} {185} (\bibinfo {year} {2007})},\ \Eprint
  {http://arxiv.org/abs/hep-ph/0610278} {arXiv:hep-ph/0610278} \BibitemShut
  {NoStop}%
\bibitem [{\citenamefont {Wu}\ and\ \citenamefont {Chen}(2019)}]{Wu:2019rog}%
  \BibitemOpen
  \bibfield  {author} {\bibinfo {author} {\bibfnamefont {Q.}~\bibnamefont
  {Wu}}\ and\ \bibinfo {author} {\bibfnamefont {D.-Y.}\ \bibnamefont {Chen}},\
  }\href {\doibase 10.1103/PhysRevD.100.114002} {\bibfield  {journal} {\bibinfo
   {journal} {Phys. Rev. D}\ }\textbf {\bibinfo {volume} {100}},\ \bibinfo
  {pages} {114002} (\bibinfo {year} {2019})},\ \Eprint
  {http://arxiv.org/abs/1906.02480} {arXiv:1906.02480 [hep-ph]} \BibitemShut
  {NoStop}%
\bibitem [{\citenamefont {Liu}\ \emph {et~al.}(2020)\citenamefont {Liu},
  \citenamefont {Yan}, \citenamefont {Ke}, \citenamefont {Li},\ and\
  \citenamefont {Xie}}]{Liu:2020orv}%
  \BibitemOpen
  \bibfield  {author} {\bibinfo {author} {\bibfnamefont {X.-H.}\ \bibnamefont
  {Liu}}, \bibinfo {author} {\bibfnamefont {M.-J.}\ \bibnamefont {Yan}},
  \bibinfo {author} {\bibfnamefont {H.-W.}\ \bibnamefont {Ke}}, \bibinfo
  {author} {\bibfnamefont {G.}~\bibnamefont {Li}}, \ and\ \bibinfo {author}
  {\bibfnamefont {J.-J.}\ \bibnamefont {Xie}},\ }\href {\doibase
  10.1140/epjc/s10052-020-08762-6} {\bibfield  {journal} {\bibinfo  {journal}
  {Eur. Phys. J. C}\ }\textbf {\bibinfo {volume} {80}},\ \bibinfo {pages}
  {1178} (\bibinfo {year} {2020})},\ \Eprint {http://arxiv.org/abs/2008.07190}
  {arXiv:2008.07190 [hep-ph]} \BibitemShut {NoStop}%
\bibitem [{\citenamefont {Liu}\ \emph {et~al.}(2022)\citenamefont {Liu},
  \citenamefont {Ling}, \citenamefont {Geng}, \citenamefont {En-Wang},\ and\
  \citenamefont {Xie}}]{Liu:2022dmm}%
  \BibitemOpen
  \bibfield  {author} {\bibinfo {author} {\bibfnamefont {M.-Z.}\ \bibnamefont
  {Liu}}, \bibinfo {author} {\bibfnamefont {X.-Z.}\ \bibnamefont {Ling}},
  \bibinfo {author} {\bibfnamefont {L.-S.}\ \bibnamefont {Geng}}, \bibinfo
  {author} {\bibnamefont {En-Wang}}, \ and\ \bibinfo {author} {\bibfnamefont
  {J.-J.}\ \bibnamefont {Xie}},\ }\href {\doibase 10.1103/PhysRevD.106.114011}
  {\bibfield  {journal} {\bibinfo  {journal} {Phys. Rev. D}\ }\textbf {\bibinfo
  {volume} {106}},\ \bibinfo {pages} {114011} (\bibinfo {year} {2022})},\
  \Eprint {http://arxiv.org/abs/2209.01103} {arXiv:2209.01103 [hep-ph]}
  \BibitemShut {NoStop}%
\bibitem [{\citenamefont {Chau}\ and\ \citenamefont
  {Cheng}(1987)}]{Chau:1987tk}%
  \BibitemOpen
  \bibfield  {author} {\bibinfo {author} {\bibfnamefont {L.-L.}\ \bibnamefont
  {Chau}}\ and\ \bibinfo {author} {\bibfnamefont {H.-Y.}\ \bibnamefont
  {Cheng}},\ }\href {\doibase 10.1103/PhysRevD.39.2788} {\bibfield  {journal}
  {\bibinfo  {journal} {Phys. Rev. D}\ }\textbf {\bibinfo {volume} {36}},\
  \bibinfo {pages} {137} (\bibinfo {year} {1987})},\ \bibinfo {note}
  {[Addendum: Phys.Rev.D 39, 2788--2791 (1989)]}\BibitemShut {NoStop}%
\bibitem [{\citenamefont {Ali}\ \emph {et~al.}(1998)\citenamefont {Ali},
  \citenamefont {Kramer},\ and\ \citenamefont {Lu}}]{Ali:1998eb}%
  \BibitemOpen
  \bibfield  {author} {\bibinfo {author} {\bibfnamefont {A.}~\bibnamefont
  {Ali}}, \bibinfo {author} {\bibfnamefont {G.}~\bibnamefont {Kramer}}, \ and\
  \bibinfo {author} {\bibfnamefont {C.-D.}\ \bibnamefont {Lu}},\ }\href
  {\doibase 10.1103/PhysRevD.58.094009} {\bibfield  {journal} {\bibinfo
  {journal} {Phys. Rev. D}\ }\textbf {\bibinfo {volume} {58}},\ \bibinfo
  {pages} {094009} (\bibinfo {year} {1998})},\ \Eprint
  {http://arxiv.org/abs/hep-ph/9804363} {arXiv:hep-ph/9804363} \BibitemShut
  {NoStop}%
\bibitem [{\citenamefont {Ali}\ \emph {et~al.}(2007)\citenamefont {Ali},
  \citenamefont {Kramer}, \citenamefont {Li}, \citenamefont {Lu}, \citenamefont
  {Shen}, \citenamefont {Wang},\ and\ \citenamefont {Wang}}]{Ali:2007ff}%
  \BibitemOpen
  \bibfield  {author} {\bibinfo {author} {\bibfnamefont {A.}~\bibnamefont
  {Ali}}, \bibinfo {author} {\bibfnamefont {G.}~\bibnamefont {Kramer}},
  \bibinfo {author} {\bibfnamefont {Y.}~\bibnamefont {Li}}, \bibinfo {author}
  {\bibfnamefont {C.-D.}\ \bibnamefont {Lu}}, \bibinfo {author} {\bibfnamefont
  {Y.-L.}\ \bibnamefont {Shen}}, \bibinfo {author} {\bibfnamefont
  {W.}~\bibnamefont {Wang}}, \ and\ \bibinfo {author} {\bibfnamefont {Y.-M.}\
  \bibnamefont {Wang}},\ }\href {\doibase 10.1103/PhysRevD.76.074018}
  {\bibfield  {journal} {\bibinfo  {journal} {Phys. Rev. D}\ }\textbf {\bibinfo
  {volume} {76}},\ \bibinfo {pages} {074018} (\bibinfo {year} {2007})},\
  \Eprint {http://arxiv.org/abs/hep-ph/0703162} {arXiv:hep-ph/0703162}
  \BibitemShut {NoStop}%
\bibitem [{\citenamefont {Li}\ \emph {et~al.}(2012)\citenamefont {Li},
  \citenamefont {Lu},\ and\ \citenamefont {Yu}}]{Li:2012cfa}%
  \BibitemOpen
  \bibfield  {author} {\bibinfo {author} {\bibfnamefont {H.-n.}\ \bibnamefont
  {Li}}, \bibinfo {author} {\bibfnamefont {C.-D.}\ \bibnamefont {Lu}}, \ and\
  \bibinfo {author} {\bibfnamefont {F.-S.}\ \bibnamefont {Yu}},\ }\href
  {\doibase 10.1103/PhysRevD.86.036012} {\bibfield  {journal} {\bibinfo
  {journal} {Phys. Rev. D}\ }\textbf {\bibinfo {volume} {86}},\ \bibinfo
  {pages} {036012} (\bibinfo {year} {2012})},\ \Eprint
  {http://arxiv.org/abs/1203.3120} {arXiv:1203.3120 [hep-ph]} \BibitemShut
  {NoStop}%
\bibitem [{\citenamefont {Bauer}\ \emph {et~al.}(1987)\citenamefont {Bauer},
  \citenamefont {Stech},\ and\ \citenamefont {Wirbel}}]{Bauer:1986bm}%
  \BibitemOpen
  \bibfield  {author} {\bibinfo {author} {\bibfnamefont {M.}~\bibnamefont
  {Bauer}}, \bibinfo {author} {\bibfnamefont {B.}~\bibnamefont {Stech}}, \ and\
  \bibinfo {author} {\bibfnamefont {M.}~\bibnamefont {Wirbel}},\ }\href
  {\doibase 10.1007/BF01561122} {\bibfield  {journal} {\bibinfo  {journal} {Z.
  Phys. C}\ }\textbf {\bibinfo {volume} {34}},\ \bibinfo {pages} {103}
  (\bibinfo {year} {1987})}\BibitemShut {NoStop}%
\bibitem [{\citenamefont {Cheng}\ and\ \citenamefont
  {Tseng}(1993)}]{Cheng:1993gf}%
  \BibitemOpen
  \bibfield  {author} {\bibinfo {author} {\bibfnamefont {H.-Y.}\ \bibnamefont
  {Cheng}}\ and\ \bibinfo {author} {\bibfnamefont {B.}~\bibnamefont {Tseng}},\
  }\href {\doibase 10.1103/PhysRevD.48.4188} {\bibfield  {journal} {\bibinfo
  {journal} {Phys. Rev. D}\ }\textbf {\bibinfo {volume} {48}},\ \bibinfo
  {pages} {4188} (\bibinfo {year} {1993})},\ \Eprint
  {http://arxiv.org/abs/hep-ph/9304286} {arXiv:hep-ph/9304286} \BibitemShut
  {NoStop}%
\bibitem [{\citenamefont {Cheng}\ and\ \citenamefont
  {Chiang}(2010)}]{Cheng:2010ry}%
  \BibitemOpen
  \bibfield  {author} {\bibinfo {author} {\bibfnamefont {H.-Y.}\ \bibnamefont
  {Cheng}}\ and\ \bibinfo {author} {\bibfnamefont {C.-W.}\ \bibnamefont
  {Chiang}},\ }\href {\doibase 10.1103/PhysRevD.81.074021} {\bibfield
  {journal} {\bibinfo  {journal} {Phys. Rev. D}\ }\textbf {\bibinfo {volume}
  {81}},\ \bibinfo {pages} {074021} (\bibinfo {year} {2010})},\ \Eprint
  {http://arxiv.org/abs/1001.0987} {arXiv:1001.0987 [hep-ph]} \BibitemShut
  {NoStop}%
\bibitem [{\citenamefont {Lu}\ \emph {et~al.}(2009)\citenamefont {Lu},
  \citenamefont {Wang}, \citenamefont {Zou}, \citenamefont {Ali},\ and\
  \citenamefont {Kramer}}]{Lu:2009cm}%
  \BibitemOpen
  \bibfield  {author} {\bibinfo {author} {\bibfnamefont {C.-D.}\ \bibnamefont
  {Lu}}, \bibinfo {author} {\bibfnamefont {Y.-M.}\ \bibnamefont {Wang}},
  \bibinfo {author} {\bibfnamefont {H.}~\bibnamefont {Zou}}, \bibinfo {author}
  {\bibfnamefont {A.}~\bibnamefont {Ali}}, \ and\ \bibinfo {author}
  {\bibfnamefont {G.}~\bibnamefont {Kramer}},\ }\href {\doibase
  10.1103/PhysRevD.80.034011} {\bibfield  {journal} {\bibinfo  {journal} {Phys.
  Rev. D}\ }\textbf {\bibinfo {volume} {80}},\ \bibinfo {pages} {034011}
  (\bibinfo {year} {2009})},\ \Eprint {http://arxiv.org/abs/0906.1479}
  {arXiv:0906.1479 [hep-ph]} \BibitemShut {NoStop}%
\bibitem [{\citenamefont {Verma}(2012)}]{Verma:2011yw}%
  \BibitemOpen
  \bibfield  {author} {\bibinfo {author} {\bibfnamefont {R.~C.}\ \bibnamefont
  {Verma}},\ }\href {\doibase 10.1088/0954-3899/39/2/025005} {\bibfield
  {journal} {\bibinfo  {journal} {J. Phys. G}\ }\textbf {\bibinfo {volume}
  {39}},\ \bibinfo {pages} {025005} (\bibinfo {year} {2012})},\ \Eprint
  {http://arxiv.org/abs/1103.2973} {arXiv:1103.2973 [hep-ph]} \BibitemShut
  {NoStop}%
\bibitem [{\citenamefont {Aoki}\ \emph {et~al.}(2020)\citenamefont {Aoki} \emph
  {et~al.}}]{FlavourLatticeAveragingGroup:2019iem}%
  \BibitemOpen
  \bibfield  {author} {\bibinfo {author} {\bibfnamefont {S.}~\bibnamefont
  {Aoki}} \emph {et~al.} (\bibinfo {collaboration} {Flavour Lattice Averaging
  Group}),\ }\href {\doibase 10.1140/epjc/s10052-019-7354-7} {\bibfield
  {journal} {\bibinfo  {journal} {Eur. Phys. J. C}\ }\textbf {\bibinfo {volume}
  {80}},\ \bibinfo {pages} {113} (\bibinfo {year} {2020})},\ \Eprint
  {http://arxiv.org/abs/1902.08191} {arXiv:1902.08191 [hep-lat]} \BibitemShut
  {NoStop}%
\bibitem [{\citenamefont {Li}\ \emph {et~al.}(2017)\citenamefont {Li},
  \citenamefont {Maris},\ and\ \citenamefont {Vary}}]{Li:2017mlw}%
  \BibitemOpen
  \bibfield  {author} {\bibinfo {author} {\bibfnamefont {Y.}~\bibnamefont
  {Li}}, \bibinfo {author} {\bibfnamefont {P.}~\bibnamefont {Maris}}, \ and\
  \bibinfo {author} {\bibfnamefont {J.~P.}\ \bibnamefont {Vary}},\ }\href
  {\doibase 10.1103/PhysRevD.96.016022} {\bibfield  {journal} {\bibinfo
  {journal} {Phys. Rev. D}\ }\textbf {\bibinfo {volume} {96}},\ \bibinfo
  {pages} {016022} (\bibinfo {year} {2017})},\ \Eprint
  {http://arxiv.org/abs/1704.06968} {arXiv:1704.06968 [hep-ph]} \BibitemShut
  {NoStop}%
\bibitem [{\citenamefont {Azevedo}\ and\ \citenamefont
  {Nielsen}(2004)}]{Azevedo:2003qh}%
  \BibitemOpen
  \bibfield  {author} {\bibinfo {author} {\bibfnamefont {R.~S.}\ \bibnamefont
  {Azevedo}}\ and\ \bibinfo {author} {\bibfnamefont {M.}~\bibnamefont
  {Nielsen}},\ }\href {\doibase 10.1103/PhysRevC.69.035201} {\bibfield
  {journal} {\bibinfo  {journal} {Phys. Rev. C}\ }\textbf {\bibinfo {volume}
  {69}},\ \bibinfo {pages} {035201} (\bibinfo {year} {2004})},\ \Eprint
  {http://arxiv.org/abs/nucl-th/0310061} {arXiv:nucl-th/0310061} \BibitemShut
  {NoStop}%
\bibitem [{\citenamefont {Wu}\ \emph {et~al.}(2024)\citenamefont {Wu},
  \citenamefont {Liu},\ and\ \citenamefont {Geng}}]{Wu:2023rrp}%
  \BibitemOpen
  \bibfield  {author} {\bibinfo {author} {\bibfnamefont {Q.}~\bibnamefont
  {Wu}}, \bibinfo {author} {\bibfnamefont {M.-Z.}\ \bibnamefont {Liu}}, \ and\
  \bibinfo {author} {\bibfnamefont {L.-S.}\ \bibnamefont {Geng}},\ }\href
  {\doibase 10.1140/epjc/s10052-024-12501-6} {\bibfield  {journal} {\bibinfo
  {journal} {Eur. Phys. J. C}\ }\textbf {\bibinfo {volume} {84}},\ \bibinfo
  {pages} {147} (\bibinfo {year} {2024})},\ \Eprint
  {http://arxiv.org/abs/2304.05269} {arXiv:2304.05269 [hep-ph]} \BibitemShut
  {NoStop}%
\bibitem [{\citenamefont {He}\ and\ \citenamefont {Chen}(2019)}]{He:2019rva}%
  \BibitemOpen
  \bibfield  {author} {\bibinfo {author} {\bibfnamefont {J.}~\bibnamefont
  {He}}\ and\ \bibinfo {author} {\bibfnamefont {D.-Y.}\ \bibnamefont {Chen}},\
  }\href {\doibase 10.1140/epjc/s10052-019-7419-7} {\bibfield  {journal}
  {\bibinfo  {journal} {Eur. Phys. J. C}\ }\textbf {\bibinfo {volume} {79}},\
  \bibinfo {pages} {887} (\bibinfo {year} {2019})},\ \Eprint
  {http://arxiv.org/abs/1909.05681} {arXiv:1909.05681 [hep-ph]} \BibitemShut
  {NoStop}%
\bibitem [{\citenamefont {Wu}\ \emph {et~al.}(2021)\citenamefont {Wu},
  \citenamefont {Chen},\ and\ \citenamefont {Matsuki}}]{Wu:2021udi}%
  \BibitemOpen
  \bibfield  {author} {\bibinfo {author} {\bibfnamefont {Q.}~\bibnamefont
  {Wu}}, \bibinfo {author} {\bibfnamefont {D.-Y.}\ \bibnamefont {Chen}}, \ and\
  \bibinfo {author} {\bibfnamefont {T.}~\bibnamefont {Matsuki}},\ }\href
  {\doibase 10.1140/epjc/s10052-021-08984-2} {\bibfield  {journal} {\bibinfo
  {journal} {Eur. Phys. J. C}\ }\textbf {\bibinfo {volume} {81}},\ \bibinfo
  {pages} {193} (\bibinfo {year} {2021})},\ \Eprint
  {http://arxiv.org/abs/2102.08637} {arXiv:2102.08637 [hep-ph]} \BibitemShut
  {NoStop}%
\bibitem [{\citenamefont {Aaij}\ \emph {et~al.}(2013)\citenamefont {Aaij} \emph
  {et~al.}}]{LHCb:2013ywr}%
  \BibitemOpen
  \bibfield  {author} {\bibinfo {author} {\bibfnamefont {R.}~\bibnamefont
  {Aaij}} \emph {et~al.} (\bibinfo {collaboration} {LHCb}),\ }\href {\doibase
  10.1103/PhysRevLett.111.112003} {\bibfield  {journal} {\bibinfo  {journal}
  {Phys. Rev. Lett.}\ }\textbf {\bibinfo {volume} {111}},\ \bibinfo {pages}
  {112003} (\bibinfo {year} {2013})},\ \Eprint {http://arxiv.org/abs/1307.7595}
  {arXiv:1307.7595 [hep-ex]} \BibitemShut {NoStop}%
\bibitem [{\citenamefont {Liu}\ \emph {et~al.}(2024{\natexlab{b}})\citenamefont
  {Liu}, \citenamefont {Ling},\ and\ \citenamefont {Geng}}]{Liu:2023cwk}%
  \BibitemOpen
  \bibfield  {author} {\bibinfo {author} {\bibfnamefont {M.-Z.}\ \bibnamefont
  {Liu}}, \bibinfo {author} {\bibfnamefont {X.-Z.}\ \bibnamefont {Ling}}, \
  and\ \bibinfo {author} {\bibfnamefont {L.-S.}\ \bibnamefont {Geng}},\ }\href
  {\doibase 10.1103/PhysRevD.109.056014} {\bibfield  {journal} {\bibinfo
  {journal} {Phys. Rev. D}\ }\textbf {\bibinfo {volume} {109}},\ \bibinfo
  {pages} {056014} (\bibinfo {year} {2024}{\natexlab{b}})},\ \Eprint
  {http://arxiv.org/abs/2312.01433} {arXiv:2312.01433 [hep-ph]} \BibitemShut
  {NoStop}%
\bibitem [{\citenamefont {Gao}\ \emph {et~al.}(2017)\citenamefont {Gao},
  \citenamefont {Shen},\ and\ \citenamefont {Yuan}}]{Gao:2017sqa}%
  \BibitemOpen
  \bibfield  {author} {\bibinfo {author} {\bibfnamefont {X.~Y.}\ \bibnamefont
  {Gao}}, \bibinfo {author} {\bibfnamefont {C.~P.}\ \bibnamefont {Shen}}, \
  and\ \bibinfo {author} {\bibfnamefont {C.~Z.}\ \bibnamefont {Yuan}},\ }\href
  {\doibase 10.1103/PhysRevD.95.092007} {\bibfield  {journal} {\bibinfo
  {journal} {Phys. Rev. D}\ }\textbf {\bibinfo {volume} {95}},\ \bibinfo
  {pages} {092007} (\bibinfo {year} {2017})},\ \Eprint
  {http://arxiv.org/abs/1703.10351} {arXiv:1703.10351 [hep-ex]} \BibitemShut
  {NoStop}%
\bibitem [{\citenamefont {Zhang}\ \emph {et~al.}(2018)\citenamefont {Zhang},
  \citenamefont {Yuan},\ and\ \citenamefont {Wang}}]{Zhang:2018zog}%
  \BibitemOpen
  \bibfield  {author} {\bibinfo {author} {\bibfnamefont {J.}~\bibnamefont
  {Zhang}}, \bibinfo {author} {\bibfnamefont {L.}~\bibnamefont {Yuan}}, \ and\
  \bibinfo {author} {\bibfnamefont {R.}~\bibnamefont {Wang}},\ }\href {\doibase
  10.1155/2018/5428734} {\bibfield  {journal} {\bibinfo  {journal} {Adv. High
  Energy Phys.}\ }\textbf {\bibinfo {volume} {2018}},\ \bibinfo {pages}
  {5428734} (\bibinfo {year} {2018})},\ \Eprint
  {http://arxiv.org/abs/1805.03565} {arXiv:1805.03565 [hep-ph]} \BibitemShut
  {NoStop}%
\bibitem [{\citenamefont {Korchin}\ and\ \citenamefont
  {Kovalchuk}(2011)}]{Korchin:2011ze}%
  \BibitemOpen
  \bibfield  {author} {\bibinfo {author} {\bibfnamefont {A.~Y.}\ \bibnamefont
  {Korchin}}\ and\ \bibinfo {author} {\bibfnamefont {V.~A.}\ \bibnamefont
  {Kovalchuk}},\ }\href@noop {} {\  (\bibinfo {year} {2011})},\ \Eprint
  {http://arxiv.org/abs/1111.4093} {arXiv:1111.4093 [hep-ph]} \BibitemShut
  {NoStop}%
\bibitem [{\citenamefont {Shi}\ \emph {et~al.}(2024)\citenamefont {Shi},
  \citenamefont {Baru}, \citenamefont {Guo}, \citenamefont {Hanhart},\ and\
  \citenamefont {Nefediev}}]{Shi:2023ntq}%
  \BibitemOpen
  \bibfield  {author} {\bibinfo {author} {\bibfnamefont {P.-P.}\ \bibnamefont
  {Shi}}, \bibinfo {author} {\bibfnamefont {V.}~\bibnamefont {Baru}}, \bibinfo
  {author} {\bibfnamefont {F.-K.}\ \bibnamefont {Guo}}, \bibinfo {author}
  {\bibfnamefont {C.}~\bibnamefont {Hanhart}}, \ and\ \bibinfo {author}
  {\bibfnamefont {A.}~\bibnamefont {Nefediev}},\ }\href {\doibase
  10.1088/0256-307X/41/3/031301} {\bibfield  {journal} {\bibinfo  {journal}
  {Chin. Phys. Lett.}\ }\textbf {\bibinfo {volume} {41}},\ \bibinfo {pages}
  {031301} (\bibinfo {year} {2024})},\ \Eprint
  {http://arxiv.org/abs/2312.05389} {arXiv:2312.05389 [hep-ph]} \BibitemShut
  {NoStop}%
\end{thebibliography}%
\end{document}